\documentclass[aps,amsmath,amssymb,showpacs,showkeys]{revtex4}
\usepackage[dvips]{graphicx,color}
\usepackage{times}
\usepackage{xcolor, soul}
\sethlcolor{yellow}
\usepackage[
  colorlinks=true,
  urlcolor=blue,
  linkcolor=red,
  citecolor=blue
]{hyperref}

\begin{document}
\title{Bianchi Type I model of universe with customized scale factors}

\author{Pranjal Sarmah}
\email[E-mail:]{p.sarmah97@gmail.com}


\author{Umananda Dev Goswami}
\email[E-mail:]{umananda2@gmail.com}

\affiliation{Department of Physics, Dibrugarh University, Dibrugarh 786004, 
Assam, India}

\begin{abstract}
According to standard cosmology, the universe is homogeneous and isotropic 
at large scales. However, some anisotropies can be observed at the local scale 
in the universe through various ways. Here we have studied the Bianchi 
type I model with customizing the scale factors to understand the anisotropic 
nature of the universe. We have considered two cases with slight modifications
of scale factors in different directions in the generalized Bianchi Type I 
metric equation, and compared the results with the $\Lambda$CDM model and also 
with available cosmological observational data. Through this study, we also want
to predict the possible degree of anisotropy present in the  early universe and
its evolution to current time by calculating the value of density parameter for anisotropy ($\Omega_{\sigma}$) for both low and high redshift ($z$) along with 
the possible relative anisotropy that exist among different directions. It is 
found that there was a relatively higher amount of anisotropy in the 
early universe and the anisotropic nature of the universe vanishes at the near 
past and the present epochs. Thus at near past and present stages of the 
universe there is no effective distinction between this anisotropic model and 
the standard $\Lambda$CDM model.   
\end{abstract}

\pacs{}
\keywords{Bianchi type I universe; anisotrpy; relative shear anisotropy; 
density parameter for anisotropy}

\maketitle
\section{Introduction}
The standard cosmology has assumed that the universe is exactly homogeneous 
and isotropic at large scales \cite{Tedesco_2018}. The spacetime of 
this kind of universe is described by the Friedmann-Lema\^itre-Robertson-Walker
 (FLRW) metric, and it is assumed to be occupied by some ideal fluids with a 
diagonal energy momentum tensor \cite{Pebbles_1994,J.Colin_2019}. Various 
cosmological observational data like, data of Type Ia Supernovae 
(SNe Ia) \cite{Hossienkhani_2021,Campanelli_2011,Akarsu_2019,Ma_2011,
Tedesco_2018,Maurya_2018, Bamba_2012}, Cosmic Microwave Background (CMB) \cite{Akarsu_2019,Bunn_1996,Costa_2004,
Hossienkhani_2021,Ma_2011,Tedesco_2018,Campanelli_2011,Copi_2004, Eriksen_2004_a,Eriksen_2004_b}, Baryon Acoustic 
Oscillation (BAO) \cite{Akarsu_2019,Hossienkhani_2021,Ma_2011,Tedesco_2018,
Campanelli_2011}, Hubble parameter \cite{Hossienkhani_2021,Campanelli_2011,
Akarsu_2019,Ma_2011,Tedesco_2018,Maurya_2018,Frieman_2008,
Perivolaropoulos_2014,Cea_2022}, the Wilkinson Microwave Anisotropy Probe (WMAP) 
\cite{Hossienkhani_2021,Frieman_2008,Perivolaropoulos_2014,Buniy_2006, Campanelli_2006, Campanelli_2007,Berera_2004, Vielva_2004, Hansen_2004} and the Large 
Scale Structure (LSS) \cite{Hossienkhani_2021,Frieman_2008} etc.\ have 
suggested that the current expansion of the  universe is accelerating 
\cite{Hossienkhani_2021,Ma_2011}. This current accelerated expansion of the 
universe is considered to be due to the presence of a large negative pressure 
component in the universe, which is referred as the Dark Energy (DE). On the 
other hand, an unknown form of matter known as the Dark Matter (DM) is thought 
to be responsible for the formation of LSS in the universe \cite{J.Colin_2019, 
Hossienkhani_2018}. However, at this point it would be appropriate to mention 
that according to the modified gravity theories scenario, the current 
accelerated expansion of the universe and DM are the manifestations of the 
modifications of spacetime behaviours at large scales, which could not be 
accounted for by the general theory of relativity \cite{Nojiri_2017,Gogoi_2021,
Parbin_2021}. Although the distribution of galaxies under the background of 
dominated DM is observed to be inhomogeneous \cite{J.Colin_2019}, other 
observational sources like different galaxy redshift catalogues have suggested 
a statistical transition from the inhomogeneity to homogeneity on scale 
exceeding $\sim$ 100 Mpc \cite{J.Colin_2019, Hogg_2005,Scrimgeour_2012}. Thus, 
the assumption of isotropy and homogeneity of the universe is applicable only 
for the large scales \cite{Tedesco_2018}. However, the standard cosmological 
models are challenged by few other puzzling cosmological observations 
\cite{Perivolaropoulos_2014,Wang_2018}, such as power asymmetry of the CMB 
perturbation maps \citep{Eriksen H. K_2007,Hoftuft_2009,Paci_2010,
Mariano_2013,Zhao_2016}, anisotropy in accelerating 
expansion rate \cite{Antoniou_2010, Mariano_2012, Wang_2014}, the large scale 
velocity flow \cite{Kashlinsky_2010,Kashlinsky_2008,Watkins_2009,Feldman_2010,
Lavaux_2010}, spatial dependence of the value of the fine structure constant
\cite{Perivolaropoulos_2014, Moss_2011, Webb_2011, King_2012, Pinho_2016},  
etc. Based on the observations of these cosmic anomalies 
\cite{Perivolaropoulos_2014}, the Planck collaboration had rightly stated 
that \textit{``The Universe is still weird and interesting"} \cite{Zhao_2016}. 
The origin of these anomalies are still unknown. These may either be arised due 
to the large statistical fluctuations or may have some physical origin like 
geometry or energy of the universe \cite{Perivolaropoulos_2014} or may favour 
of a preferred cosmological direction, and perhaps depict the anisotropic nature
of the universe.

Testing the large-scale geometry of the universe by using cosmological scale 
is one of the major challenges of modern cosmology \cite{Campanelli_2011}. 
Reliable observational data from the various sources like WMAP, BAO, Planck 
\cite{Planck_2018} etc.\ have given new insight on the study of the basic 
assumptions (i.e.\ isotropy and homogeneity) of the standard cosmological 
models and also shown the anisotropic characteristics of the universe at least 
at the local scale. In our real universe, there exist some kind of strange 
motions due to the local 
inhomogeneity and anisotropy of surrounding structures, which can not be 
neglected \cite{J.Colin_2019}. To explain the anisotropic character of the 
universe, spatially homogeneous and anisotropic cosmological models play a 
significant role to describe the large-scale behaviour of the universe. These 
models have been widely studied under the regime of general relativity to 
understand the relativistic picture of the universe at its early stage 
\cite{Maurya_2018}. Thus, we need a simple and convincing cosmological model 
with a relativistic background metric \cite{Akarsu_2019}, which allows 
directional scale factors \cite{Akarsu_2019,Tedesco_2018,Campanelli_2011, 
Maurya_2018,Paul_2008} while maintaining the spatial homogeneity and flatness. 
The Bianchi Type I model \cite{Akarsu_2019,Tedesco_2018,Maurya_2018,
Campanelli_2011, Hossienkhani_2018,Paul_2008,Perivolaropoulos_2014} and its 
metric fulfill the above criteria and hence it is suitable for study the 
anisotropic nature of the universe in its early stages as well as in the 
current scenario. Bianchi Type I model describes the anisotropic nature of the 
universe by considering the anisotropic and homogeneous background 
\cite{Tedesco_2018}. It gives a very small deviation from the exact isotropy. 
Hence, use of Bianchi Type I geometry corresponds to replacing the spatially 
flat FLRW background by Bianchi Type I background metric \cite{Tedesco_2018}. 
Therefore, the Bianchi Type I geometry can be considered as an alternative 
option to the FLRW metric to study the various cosmic anomalies as mentioned 
above. 

Keeping in view of the above points, in this study, we have used the Bianchi 
Type I metric \citep{Akarsu_2019,Maurya_2018,Hossienkhani_2018,Paul_2008,
Perivolaropoulos_2014} with the ``customized scale factors" in different 
directions. Here we have considered two cases. For case I, we have modified 
directional scale factors in the Bianchi type I metric by considering the 
second and third scale factors in terms of the first one with some 
multiplicative constants. For case II, we have again considered the second and 
third scale factors in terms of the first one, but here we have added constant 
terms with them rather than multiplying. With these two cases of scale 
factors in the metric, we have derived various cosmological parameters 
\cite{Tedesco_2018,Akarsu_2019,Maurya_2018,Perivolaropoulos_2014,Frieman_2008,
Hossienkhani_2018,Gogoi_2021,Paul_2008,Barrow_1997} including the directional 
Hubble parameters, average or mean Hubble parameter, deceleration parameter, 
shear scalar, relative shear anisotropy parameter or Hubble normalized shear 
parameter, average anisotropic expansion, density parameters (matter, 
radiation, anisotropy and dark energy), distance modulus, Equation of State 
(EoS) etc. We have also studied the evolution of various cosmological 
parameters with respect to cosmological redshift ($z$) \cite{Frieman_2008,
Gogoi_2021,Hossienkhani_2018} by plotting those parameters against $z$ or 
($1+z$) and compared them with the standard cosmological model ($\Lambda$CDM 
model) and also with the available observational data of various sources for 
both the cases.
 
This paper is organized as follows. In Section \ref{2}, the basics of the 
Bianchi Type I model have been discussed. Here, we have mainly included the 
various mathematical equations and expressions of the Bianchi type I universe 
from various literatures. In Section \ref{Model}, we have derived all the 
mathematical equations and expressions using the ``customized scale factors 
approach" for the Bianchi Type I model of the universe. This section is 
subdivided into two subsections \ref{s1} and \ref{s2}, in which we have 
derived all the expressions and equations for two separate cases respectively. 
In Section \ref{4}, we have made graphical analysis of the various cosmological
parameters with respect to $z$ and tried to give the explanations of those 
results obtained from the plots for both the cases. The paper has been 
summarized in Section \ref{5} with conclusions. Throughout this work we use the
geometrized unit system, where $c = G = 1$ with $(-,+,+,+)$ metric convention. 

\section{Basic equations of Bianchi type I model}\label{2}
The general form of the metric for the Bianchi Type I universe \cite{Perivolaropoulos_2014,Akarsu_2019,Paul_2008,Maurya_2018,Hossienkhani_2018} can be 
written as
\begin{equation}\label{Bianchi_Metric}
ds^2= -\,dt^2+\sum_{i=1}^{3}a_{i}(dx^i)^{2},
\end{equation}
where $a_{i}$ is the directional scale factor along the $i$th direction, 
which is the function of time $t$ only. The average expansion scale factor 
$(a_{1}a_{2}a_{3})^{\frac{1}{3}}$ arises from the average Hubble parameter 
defined as $$H=\frac{1}{3} {\sum\limits_{i=1}^{3}H_{i}}$$ with 
$H_{i}=\frac{\dot{a_{i}}}{a_{i}}$ is the directional Hubble parameter along 
the $i$th direction 
The most general form of the energy-momentum tensor $T_{\mu\nu}$ for the given 
metric can be considered in the form:
\begin{equation}\label{Stress_energy_tensor}
 T_{\mu}^{\nu}= diag\big[-\rho,P,P,P\big], 
\end{equation}
where $\rho$ is the energy density and $P$ is the isotropic pressure of a 
perfect fluid. Now, in view of the present scenario of our accelerating 
universe, we consider the Einstein field equations with the cosmological 
constant $\Lambda$ as
\begin{equation}\label{field_equation}
R_{\mu\nu}-\frac{1}{2}g_{\mu\nu}R+\Lambda g_{\mu\nu}=8\pi T_{\mu\nu},
\end{equation}
where $R_{\mu\nu}$ is the Ricci tensor, R is the Ricci scalar and $g_{\mu\nu}$ 
is the metric tensor. For the above metric (\ref{Bianchi_Metric}) and the 
energy-momentum tensor \eqref{Stress_energy_tensor}, the following set of 
field equations can be obtained from the Einstein field equations: 
\begin{align}\label{temporal_component}
H_{1}H_{2}+H_{1}H_{3}+H_{2}H_{3}&=8\pi \rho + \Lambda,\\[5pt]
\label{spatial_component_I}
\frac{\ddot{a_{2}}}{a_{2}}+\frac{\ddot{a_{3}}}{a_{3}}+H_{2}H_{3}&=-\,8\pi  P+\Lambda,\\[5pt]
\label{spatial_component_II}
\frac{\ddot{a_{3}}}{a_{3}}+\frac{\ddot{a_{1}}}{a_{1}}+H_{3}H_{1}&=-\,8\pi  P+\Lambda,\\[5pt]
\label{spatial_component_III}
\frac{\ddot{a_{1}}}{a_{1}}+\frac{\ddot{a_{2}}}{a_{2}}+H_{1}H_{2}&=-\,8\pi  P+\Lambda.
\end{align}
Combining Eqs.\ (\ref{spatial_component_I}), (\ref{spatial_component_II}) and 
(\ref{spatial_component_III}) and then subtracting Eq.\ 
(\ref{temporal_component}) from the combined one, we may write the resulting
Friedmann equation as
\begin{equation}\label{Friedmann_equation}
\sum_{i=1}^{3}\frac{\ddot{a_{i}}}{a_{i}}=-\,4\pi \big(3P+\rho\big)+\Lambda.
\end{equation}
The continuity equation in this model of the universe is given by 
\cite{Paul_2008},
\begin{equation}\label{continuity_equation}
\dot{\rho}+ \big(\rho+P\big) \sum_{i=1}^{3}\frac{\dot a_{i}}{a_{i}}=0.
\end{equation}
Considering the EoS $P=\omega \rho$ in Eq.\ (\ref{continuity_equation}), in 
which the parameter $\omega$ tells us about the nature of energy or matter 
density in the universe, for example, $\omega=0$ for the non-relativistic 
matter, $\omega=\frac{1}{3}$ for the radiation and $\omega=-1$ for the 
vacuum energy \cite{Frieman_2008, Maurya_2018,Akarsu_2019}, the energy 
density of the universe can be found as 
\begin{equation}\label{density}
\rho=\rho_{0}\prod_{i=1}^{3} a_{i}^{-(1+\omega)}.
\end{equation}
Here $\rho_0$ is the present energy density of the universe. From this
equation it is seen that the future energy density of this model of the
universe is the EoS parameter $\omega$ dependent, i.e.\ on the nature of 
the matter or energy content of the universe. It is clear that $\rho \propto 
(a_1 a_2 a_3)^{-1}$ for the matter, $\rho \propto  
(a_1 a_2 a_3)^{-\frac{4}{3}}$ for the radiation and $\rho = \rho_0$ (constant) 
for the vacuum energy dominated cases respectively.

Using Eq.\ (\ref{temporal_component}) the Hubble parameter of this universe
can be expressed in terms of the present density parameters of the universe as
\begin{equation}\label{Hubble_parameter}
H= H_{0}{\sqrt{\Omega_{m0}(a_{1}a_{2}a_{3})^{-1}+\Omega_{r0}(a_{1}a_{2}a_{3})^{-\frac{4}{3}}+\Omega_{\Lambda 0}+\Omega_{\sigma 0}(a_{1}a_{2}a_{3})^{-2}}},
\end{equation}
where $H_0$ is the current Hubble parameter, and $\Omega_{m0} = 
8\pi \rho_{m0}/3H_{0}^{2}$ is the density parameters for the matter content, 
$\Omega_{r0} = 8\pi  \rho_{r0}/3H_{0}^{2}$ is the density parameter for the 
radiation content, $\Omega_{\Lambda 0} = \Lambda/3H_{0}^{2}$ is the density 
parameter for the vacuum energy and $\Omega_{\sigma 0} = 
\sigma_{0}^{2}/3H_{0}^{2}$ is the density parameter for the anisotropy 
\cite{Akarsu_2019,Maurya_2018,Hossienkhani_2021} of the present universe. 
Here, $\rho_{m0}$ and $\rho_{r0}$ are current values of matter density 
$\rho_{m}$ and radiation density $\rho_{r}$ of the universe respectively. 
{The term $\sigma_{0}^{2}$ in $\Omega_{\sigma 0}$ is the current value of 
shear scalar. The shear scalar is the parameter through which the contribution 
of the expansion anisotropy in Bianchi type I model of universe can be 
quantified. It arises due to the different scale factor taken for different 
directions in the Bianchi model of universe. For the isotropic case, 
$\sigma_{0}^{2} = 0$.} The current value  { of shear scalar, 
i.e.~$\sigma_{0}^{2}$} is related with the shear scalar  {$\sigma^{2}$} by 
the relation $\sigma^{2}= \sigma_0^2(a_1 a_2 a_3 )^{-2}$. {In general} the shear scalar {$\sigma^{2}$} in terms of the average and directional Hubble parameters can be written as 
\cite{Maurya_2018,Akarsu_2019,Paul_2008}
\begin{equation}\label{shear_scalar_II}
\sigma^{2}= \frac{1}{2}\Big[\sum_{i=1}^{3}H_{i}^{2}-3H^2\Big].
\end{equation}

Again, the deceleration parameter for the Bianchi Type I universe can be 
derived as
\begin{equation}\label{dec_par_I}
q=-\frac{1}{3H^2}\bigg(\frac{\ddot a_1}{a_1}+\frac{\ddot a_2}{a_2}+\frac{\ddot a_3}{a_3}\bigg)+\frac{2\sigma^2}{3H^2}.
\end{equation}
Using Eq.\ \eqref{Friedmann_equation} and the EoS mentioned above, this equation
for the deceleration parameter $q$ can expressed in terms of density parameters
of the universe as
\begin{equation}\label{dec_par_II}
q=\frac{1}{2}\Omega_m + \Omega_r -\Omega_{\Lambda}+2\,\Omega_{\sigma},
\end{equation}
where $\Omega_m = 8\pi \rho_{m}/3H^{2}$, $\Omega_r = 8\pi \rho_r/3H^{2}$,
$\Omega_{\Lambda} = \Lambda/3H^{2}$ and $\Omega_{\sigma}=\frac{\sigma^2}{3H^2}$ \cite{Akarsu_2019} are the matter density parameter, radiation density 
parameter, vacuum energy density parameter and anisotropy or shear density 
parameter respectively of the universe at any given instant. Similarly, the 
average anisotropic expansion \cite{Maurya_2018, Paul_2008} for Bianchi type I 
universe is 
\begin{equation}\label{Avg_exp}
A=\frac{1}{3}\sum_{i=1}^{3}\Big(\,\frac{\Delta H_i}{H}\Big)^2 = 
2\,\Omega_{\sigma}, \;\; \Delta H_i= (H_i-H).
\end{equation} 

The cosmological redshift ($z$) is another cosmological parameter, which is a 
very important parameter to understand the evolution and history of the 
universe in the sense that it is a directly observable parameter, not the 
scale factor. Hence to quantify the 
effect of anisotropy in the observable domain of the universe, it is necessary 
to express all other cosmological parameters mentioned above as the functions 
of $z$. In this context, it is to be noted that in the case of the Bianchi 
type I universe, there exist three redshifts in three spatial directions due 
to the directional dependence of scale factor. If $z_1, z_2, z_3$ are these 
three redshift parameters along three spatial directions, then they can be 
defined as
\begin{equation}
\frac{a_1(t)}{a_{10}}=\frac{1}{1+z_{1}}, \;\; 
\frac{a_2(t)}{a_{20}}=\frac{1}{1+z_{2}} \;\;\mbox{and}\;\;
\frac{a_3(t)}{a_{30}}=\frac{1}{1+z_{3}}.
\end{equation}
Here, $a_{10}, a_{20}, a_{30}$ are the scale factors of present time along the 
three spatial directions. Now, from the current observations, we have taken 
$a_{10} = a_{20}= a_{30} = 1$ \cite{Maurya_2018} and thus the scale factors 
can be rewritten in terms of $z_1, z_2, z_3$ as

\begin{equation}\label{redshift_I}
a_1(t)=\frac{1}{1+z_{1}},\;\; {a_2(t)}=\frac{1}{1+z_{2}}\;\; \mbox{and}\;\;
{a_3(t)}=\frac{1}{1+z_{3}}.
\end{equation}
With this form of the scale factors, the directional Hubble parameters ($H_i)$ 
can be written in terms of $z_i$ as
\begin{equation}\label{dir_hubble}
H_i=-\,\frac{\dot{z}_i}{1+z_i}
\end{equation}
and hence the average Hubble parameter takes the form:
\begin{equation}\label{avg_hubble}
H= -\frac{1}{3}\sum_{i=1}^{3}\frac{\dot{z}_i}{1+z_i}.
\end{equation}
Thus, Eq.\ (\ref{Hubble_parameter}) for the average Hubble parameter can be 
rewritten in terms of redshift parameters $(z_1, z_2, z_3)$ as
\begin{equation}\label{H_E}
H=H_0 \sqrt{E(z_1,z_2,z_3)},
\end{equation}
\begin{multline}\label{E_z}
\!\!\!\!\!\!\!\mbox{where}\;\;
E(z_1,z_2,z_3) = \Omega_{m0}\big[(1+z_1)(1+z_2)(1+z_3)\big]\\+\Omega_{r0}\big[(1+z_1)(1+z_2)(1+z_3)\big]^{\frac{4}{3}}+\Omega_{\Lambda 0}+\Omega_{\sigma 0}\big[(1+z_1)(1+z_2)(1+z_3)\big]^{2}.
\end{multline}
With the help of Eqs.\ (\ref{dir_hubble}), (\ref{avg_hubble}) and (\ref{H_E}), 
the age of the universe can be expressed in terms of the directional redshift 
parameters $(z_1,z_2,z_3)$ as follows: 
\begin{multline}\label{age_univ}
t_{age}=\int_{0}^{t}dt=\frac{1}{3H_0}\Bigg[\int_{0}^{\infty}\frac{dz_1}{(1+z_1)\sqrt{E(z_1,z_2,z_3)}}+\int_{0}^{\infty}\frac{dz_2}{(1+z_2)\sqrt{E(z_1,z_2,z_3)}}+\int_{0}^{\infty}\frac{dz_3}{(1+z_3)\sqrt{E(z_1,z_2,z_3)}}\Bigg].
\end{multline}
Other cosmological parameters in terms of redshift can be derived for this model
of the universe as follows:\\

(i) Deceleration parameter:
\begin{multline}\label{dec_final}
q(z_1,z_2,z_3)=\frac{1}{E(z_1,z_2,z_3)}\bigg[\frac{1}{2}\,\Omega_{m0}\,(1+z_1)(1+z_2)(1+z_3) +\Omega_{r0}\Big\{(1+z_1)(1+z_2)(1+z_3)\Big\}^{\frac{4}{3}}\\-\,\Omega_{\Lambda 0}+2\,\Omega_{\sigma 0}\Big\{(1+z_1)(1+z_2)(1+z_3)\Big\}^{2}\bigg].
\end{multline}

 (ii) Equation of State:
\begin{equation}\label{eos_final}
\omega(z_1,z_2,z_3) = \frac{\frac{1}{3}\,\Omega_{r0}\big[(1+z_1)(1+z_2)(1+z_3)\big]^{\frac{4}{3}}-\Omega_{\Lambda 0}}{\Omega_{m0}\big[(1+z_1)(1+z_2)(1+z_3)\big]+\Omega_{r0}\big[(1+z_1)(1+z_2)(1+z_3)\big]^{\frac{4}{3}}+\Omega_{\Lambda 0}}.
\end{equation}

(iii) Ricci Scalar:
\begin{equation}\label{Ricci_tensor_final}
R(z_1,z_2,z_3) = 3H_0^2\Big[\Omega_{m0}(1+z_1)(1+z_2)(1+z_3)+4\,\Omega_{\Lambda 0}\Big].
\end{equation}

(iv) Luminosity distance:
\begin{multline}\label{lumin_final}
d_L(z_1,z_2,z_3) =\frac{\big[(1+z_1)(1+z_2)(1+z_3)\big]^{\frac{1}{3}}}{3H_0}\int_{0}^{\infty} \Big\{\frac{\big[(1+z_2)(1+z_3)\big]^{\frac{1}{3}}}{(1+z_1)^{\frac{2}{3}}\sqrt{E(z_1,z_2,z_3)}}\,dz_1 + \frac{\big[(1+z_1)(1+z_3)\big]^{\frac{1}{3}}}{(1+z_2)^{\frac{2}{3}}\sqrt{E(z_1,z_2,z_3)}}\,dz_2\\+ \frac{\big[(1+z_1)(1+z_2)\big]^{\frac{1}{3}}}{(1+z_3)^{\frac{2}{3}}\sqrt{E(z_1,z_2,z_3)}}\,dz_3\Big\}.
\end{multline}

(v) Distance Modulus:
\begin{equation}\label{dist_mod}
D_m = 5 \log{d_L} + 25.
\end{equation}
These are some important expressions in general Bianchi Type I cosmology in 
generic form, which will be useful to study the properties of the universe 
in terms of the observable parameter $z$ for our considered cases as discussed 
in the following sections.

\section{Customized scale factor approach for Bianchi Type I model} 
\label{Model}
In this section we implement two different cases to the Bianchi type I model 
for which we use two different types of directional scale factors to calculate 
the various cosmological parameters as discussed below.
\subsection{\underline{Case I}}\label{s1}
In this case we consider the following set of directional scale factors:
\begin{equation}\label{scale_1}
a_1=a(t), \;\; a_2=\alpha\, a(t), \;\; \mbox{and}\;\; a_3=\beta\, a(t),
\end{equation}
where $\alpha$ and $\beta$ are two multiplicative constants. Since $\alpha$ and
$\beta$ are two time independent constants, the directional dependence of scale
factors or the anisotropy in the universe in this case should be considerable 
only for very small values of the scale factor $a(t)$, i.e.\ for the 
cosmological redshift $z\gg 1$. {The major advantage of these forms of 
scale factors is that we can express every equation and cosmological parameter 
in terms of a single scale factor multiplied by some constants. It helps to 
simplify the problem and hence reduces the complexity.} The directional Hubble parameters for the above considered
set of scale factors take the same form as in the case of the isotropic 
situation, i.e.
\begin{equation}\label{dir_hub_case_I}
H_1=H_2=H_3=\frac{\dot a}{a},
\end{equation} 
and consequently the average Hubble parameter is
\begin{equation}\label{avg_hubble_case_I}
H=\frac{\dot a}{a}.
\end{equation}
Also it can be easily seen that
\begin{equation}
\frac{\ddot a_1}{a_1}=\frac{\ddot a_2}{a_2}=\frac{\ddot a_3}{a_3}=\frac{\ddot a}{a}.
\end{equation} 
Thus the temporal component of the Einstein field equations, i.e.\ Eq.\ 
(\ref{temporal_component}) is transformed for this case as
\begin{equation}\label{temporal_case_I}
H^2=\frac{8\pi  \rho}{3}+\frac{\Lambda}{3}.
\end{equation}
Whereas all the three spatial components of the Einstein field equations, given
in Eqs.\ (\ref{spatial_component_I}), (\ref{spatial_component_II}),
(\ref{spatial_component_III}) take the same form for this case as
\begin{equation}\label{spatial_case_I}
2\,\frac{\ddot a}{a}+H^2=-\,8\pi P+\Lambda,
\end{equation}
and hence the combination of all the three spatial field equations leads to the
equation: 
\begin{equation}\label{combined_comp_case_I}
6\,\frac{\ddot a}{a}+3H^2=-\,8\pi\, (3P)+3\Lambda.
\end{equation} 
Now multiplication of Eq.\ (\ref{temporal_case_I}) by a factor $3$ and then 
subtracting it from Eq.\ (\ref{combined_comp_case_I}) give rise a new equation:
\begin{equation}\label{friedmann_case_I}
\frac{\ddot a}{a}=-\frac{4\pi\,  (\rho+3P)}{3}+\frac{\Lambda}{3}.
\end{equation}
It should be pointed out that this Eq.\ (\ref{friedmann_case_I}) can also be 
derived directly from Eq.\ (\ref{Friedmann_equation}). 
Eqs.\ (\ref{temporal_case_I}) and 
(\ref{friedmann_case_I}) are the two independent form of field equations.
Similarly the Eq.\ (\ref{continuity_equation}) i.e.\ the continuity equation 
can be rewritten in this case as
\begin{equation}\label{continuity_case_I}
\dot{\rho}+3\,{\frac{\dot a}{a}}\,(\rho+P)=0.
\end{equation}
Again Eq.\ (\ref{density}), i.e.\ the expression of energy density for this case
becomes,
\begin{equation}\label{density_case_I}
\rho=(\alpha\beta)^{-(1+\omega)}{\rho_{0}}\,{a^{-3(1+\omega)}}.
\end{equation}
It is clear that this expression is different from the isotropic situation as
it depends on the parameters $\alpha$ and $\beta$. Hence, although the 
expressions for directional and average Hubble parameters, and some other 
expressions derived above look identical to the isotropic situation, they 
would behave differently because of the different form of the energy density in 
this case. This can be seen clearly from the expression 
(\ref{Hubble_parameter}) for the Hubble parameter, which can rewritten for 
this case as
\begin{equation}\label{hub_dens_case_I}
H= H_{0}{\sqrt{(\alpha\beta)^{-1}\Omega_{m0}\,a^{-3}+(\alpha\beta)^{-\frac{4}{3}}\Omega_{r0}\,a^{-4}+\Omega_{\Lambda 0}}}.
\end{equation} 
Here $\Omega_{\sigma}=0$ as $\sigma^{2}=0$ from Eq.\ (\ref{shear_scalar_II}) 
for this case. We see that the matter and radiation parts of energy density
contributed differently in the Hubble parameter than the isotropic condition 
because of the expression \eqref{density_case_I} for the energy density. This
situation will be similar to all other cosmological parameters discussed here. 
The deceleration parameter of Eq.\ (\ref{dec_par_I}) in this case reduces to
\begin{equation}
q=-\frac{\ddot a}{aH^2}.
\end{equation} 
And in terms of density parameter, $q$ can be rewritten as
\begin{equation}\label{q_dens_case_I_}
q=\Omega_r+\frac{1}{2}\,\Omega_m-\Omega_{\Lambda}.
\end{equation} 

If we denote $z_1 = z$, the relations between the cosmological redshift 
parameters and the directional scale factors given in Eq.\ \eqref{redshift_I} 
will be transformed in this case as
\begin{equation}\label{a_Z_caseI}
a_1(t)=\frac{1}{1+z},\;\; {a_2(t)}=\frac{\alpha}{1+z},\;\;\mbox{and}\;\;
{a_3(t)}=\frac{\beta}{1+z}.
\end{equation}
Here, the other two redshift parameters $z_2$ and $z_3$ can be found as
$$z_2 = \frac{1+z}{\alpha} - 1,\;\; \mbox{and}\;\; 
z_3 = \frac{1+z}{\beta} - 1.$$   
Accordingly, the directional Hubble parameters can be written as
\begin{equation}
H_i= -\frac{\dot{z}}{1+z}.
\end{equation}
and hence the form of the expression of the average Hubble parameter becomes 
exactly the same as the directional Hubble parameters' expression given above.
Thus, in terms of the cosmological redshift parameters 
Eq.\ (\ref{hub_dens_case_I}) can be rewritten as
\begin{equation}\label{H_Z_caseI}
H= H_{0}{\sqrt{(\alpha \beta)^{-1} \Omega_{m0}(1+z)^{3}+(\alpha \beta)^{-\frac{4}{3}}\Omega_{r0}(1+z)^{4}+\Omega_{\Lambda 0}}}\;\; = H_{0}\sqrt{E(z)},
\end{equation}
where
\begin{equation}\label{E_caseI}
E(z)=(\alpha \beta)^{-1} \Omega_{m0}(1+z)^{3}+(\alpha \beta)^{-\frac{4}{3}}\Omega_{r0}(1+z)^{4}+\Omega_{\Lambda 0}.
\end{equation}
Consequently, the age of the universe given by Eq.\ (\ref{age_univ}) can be 
rewritten for this case as
\begin{equation}\label{age_caseI}
t_{age}=\int_{0}^{t}dt=\frac{1}{H_0}\int_{0}^{\infty}\frac{dz}{(1+z)\sqrt{E(z)}}.
\end{equation}
Similarly, with the same approach the other cosmological parameters for the 
present case can be obtained as a function of redshift parameters as\\ 

(i) Deceleration parameter:
\begin{equation}\label{q_Z_caseI}
q(z)=\frac{1}{\sqrt{E(z)}}\Big[\frac{1}{2}(\alpha \beta)^{-1}\Omega_{m0}(1+z)^{3}+(\alpha \beta)^{-\frac{4}{3}}\Omega_{r0}(1+z)^{4}-\Omega_{\Lambda 0}\Big].
\end{equation}

(ii) Equation of State:
\begin{equation}\label{eos_Z_caseI}
\omega(z) = \frac{\frac{1}{3}\Omega_{r0}(\alpha \beta)^{-\frac{4}{3}}(1+z)^{4}-\Omega_{\Lambda 0}}{(\alpha \beta)^{-1}\Omega_{m0}(1+z)^{3}+\Omega_{r0}(\alpha \beta)^{-\frac{4}{3}}(1+z)^{4}+\Omega_{\Lambda 0}}.
\end{equation}

(iii) Ricci Scalar:
\begin{equation}\label{R_Z_caseI}
R(z)= 3H_0^2\Big[(\alpha \beta)^{-1}\Omega_{m0}(1+z)^{3}+4\Omega_{\Lambda 0}\Big].
\end{equation}

(iv) Luminosity distance:
\begin{equation}\label{Lumin_Z_caseI}
d_L=\frac{(\alpha \beta)^{-\frac{2}{3}}(1+z)}{H_0}\int_{0}^{\infty} \frac{dz}{\sqrt{E(z)}}.
\end{equation}

(v) Density parameter of matter:
\begin{equation}\label{dens_mat_I}
\Omega_{m}(z) =  \frac{\Omega_{m0}(\alpha\beta)^{-1}(1+z)^{3}}{E(z)}.
\end{equation}

{These are the forms of expressions of the cosmological parameters we have 
obtained for the given set of scale factors in Eq.~({\ref{scale_1}}). The 
detailed analysis of these expressions are done graphically in 
section {\ref{4}}}.

\subsection{\underline{Case II}}\label{s2}

In this case we consider that scale factors of the three dimensional universe 
differ from each other by some constant amounts and hence here we consider the 
following set of anisotropic scale factors:
\begin{equation}\label{scale1_caseII}
a_1=a(t),\;\; a_2=a(t)+\delta, \;\; \mbox{and}\;\; a_3=a(t)+\gamma,
\end{equation}
where $\delta$ and $\gamma$ are two constants.  {As in the previous case, 
these form of scale factors help us to reduce the complexity and hence easier 
to deal with the problem.} For convenience, this set of scale factors can also 
be written as
\begin{equation}\label{scale1_caseII_ad}
a_1=a(t),\;\; a_2=a(t)f_{a\delta}, \;\; \mbox{and}\;\; a_3=a(t)f_{a\gamma},
\end{equation}
where $f_{a\delta} = 1+\delta/a(t)$ and $f_{a\gamma} = 1+\gamma/a(t)$. It is 
clear that $f_{a\delta}, f_{a\gamma}\rightarrow 1$, when the scale factor 
$a(t)\rightarrow \infty$ or the cosmological redshift $z\rightarrow -1$. On 
the other hand when $a(t)\rightarrow 0$ or $z\gg1$, $f_{a\delta}$, $f_{a\gamma} \gg 1$. Thus, according to this case the universe will be almost in an 
isotropic state in the distant future and the anisotropy in the universe was 
significant in the distant past as in the case I. With this set of scale 
factors \eqref{scale1_caseII_ad} the directional Hubble parameters of the 
anistropic universe becomes:
\begin{equation}\label{dir_hub1_caseII}
H_1=\frac{\dot a}{a},\;\; 
H_2=\frac{H_1}{f_{a\delta}},\;\;
H_3=\frac{H_1}{f_{a\gamma}}
\end{equation}
and hence the average Hubble parameter can be written as
\begin{equation}\label{avg_hub_caseII}
H=\frac{H_1}{3}\Big[1 + f_{a\delta}^{-1} + f_{a\gamma}^{-1} \Big].
\end{equation}
Again,
\begin{equation}
\frac{\ddot a_1}{a_1}=\frac{\ddot a}{a},\;\; 
\frac{\ddot a_2}{a_2}=\frac{\ddot a}{a}\,f_{a\delta}^{-1},\;\; 
\frac{\ddot a_3}{a_3}=\frac{\ddot a}{a}\,f_{a\gamma}^{-1}.
\end{equation}
The Einstein field equations as mentioned in Eqs.\ (\ref{temporal_component}),
(\ref{spatial_component_I}), (\ref{spatial_component_II}) and 
(\ref{spatial_component_III}) are transformed for the present case as
\begin{align}\label{temp_caseII}
H_1^2\Big[f_{a\delta}^{-1} + f_{a\gamma}^{-1} + \big(f_{a\delta}f_{a\gamma}\big)^{-1}\Big]& = 8\pi \rho+\Lambda,\\[5pt]
\label{spatial1_caseII}
\frac{\ddot{a}}{a}\Big[f_{a\delta}^{-1} + f_{a\gamma}^{-1}\Big] + H_1^2\big(f_{a\delta}f_{a\gamma}\big)^{-1} & = -\,8\pi P+\Lambda,\\[5pt]
\label{spatial2_caseII}
\frac{\ddot{a}}{a}\Big[1+ f_{a\gamma}^{-1}\Big] + H_1^2\,f_{a\gamma}^{-1} & = -\,8\pi P+\Lambda,\\[5pt]
\label{spatial3_caseII}
\frac{\ddot{a}}{a}\Big[1+ f_{a\delta}^{-1}\Big] + H_1^2\,f_{a\delta}^{-1} & = -\,8\pi P+\Lambda.
\end{align}
And Eq.\ (\ref{Friedmann_equation}) can be rewritten as
\begin{equation}\label{friedmann_caseII}
\frac{\ddot{a}}{a}\Big[1 + f_{a\delta}^{-1} + f_{a\gamma}^{-1} \Big] =-\,4\pi (3P+\rho)+\Lambda.
\end{equation}
The continuity equation for this case becomes,
\begin{equation}\label{cont_caseII}
\dot{\rho}+ \big(\rho+P\big)H_1\Big[1 + f_{a\delta}^{-1} + f_{a\gamma}^{-1}\Big] = 0
\end{equation}
and the energy density Eq.\ (\ref{density}) can be rewritten as
\begin{equation}\label{dens_caseII}
\rho=\rho_0 \big(f_{a\delta}f_{a\gamma})^{-(1+\omega)}a^{-3(1+\omega)}.
\end{equation}
Similarly, the expression of shear scalar takes the form:
\begin{equation}\label{shear_scalar_caseII}
\sigma^2=\frac{H_1^2}{3}\bigg[1 + f_{a\delta}^{-2} + f_{a\gamma}^{-2} -\Big(f_{a\delta}^{-1} + f_{a\gamma}^{-1} + \big(f_{a\delta}f_{a\gamma}\big)^{-1}\Big)\bigg].
\end{equation}
This expression of shear scalar can be used to calculate the density parameter 
for anisotropy, $\Omega_{\sigma}=\sigma^2/3H^2$. 
The Hubble parameter expression (\ref{Hubble_parameter}) for this case can be 
found as
\begin{equation}\label{hubble_density_caseII}
H=H_0\sqrt{\Omega_{mo}\big(f_{a\delta}f_{a\gamma}\big)^{-1}a^{-3} +\Omega_{ro}\big(f_{a\delta}f_{a\gamma}\big)^{-\frac{4}{3}} a^{-4}+\Omega_{\Lambda 0}+\Omega_{\sigma 0}\big(f_{a\delta}f_{a\gamma}\big)^{-2}a^{-6}}
\end{equation}
The deceleration parameter of Eq.\ (\ref{dec_par_I}) for the present case 
takes the form:
\begin{equation}\label{q_caseII}
q=-\frac{1}{3H^2}\frac{\ddot a}{a}\Big[1+f_{a\delta}^{-1} + f_{a\gamma}^{-1}\Big]+2\,\Omega_{\sigma}.
\end{equation}
However, the expression of $q$ in terms of density parameters is same as 
Eq.\ (\ref{dec_par_II}).

As in the case I, in this case also all cosmological parameters need to be 
expressed in terms of the cosmological redshift parameter, which is an 
observable quantity. The relation between cosmological redshift parameters and 
directional scale factors in this present case will take the form:
\begin{equation}\label{a_Z_caseII}
a_1(t)=\frac{1}{1+z},\;\; {a_2(t)}=\frac{1}{1+z}+\delta,\;\; \mbox{and}\;\;
{a_3(t)}={\frac{1}{1+z}}+\gamma.
\end{equation}
As earlier, for the convenient notational purpose this set of scale factors can
 also be re-expressed as
\begin{equation}\label{a_Z_caseIIad}
a_1(t) = \frac{1}{1+z},\;\; {a_2(t)}=\frac{f_{z\delta}}{1+z},\;\; \mbox{and}\;\;
{a_3(t)}={\frac{f_{z\gamma}}{1+z}},
\end{equation}
where $f_{z\delta} = 1 + (1+z)\delta$ and $f_{z\gamma} = 1 + (1+z)\gamma$. 
Indeed, the behaviours of $f_{z\delta}$ and $f_{z\gamma}$ are exactly same 
as $f_{a\delta}$ and $f_{a\gamma}$ for different values of $z$ as discussed
above. Here the directional cosmological redshift parameters $z_2$ and $z_3$ 
can be obtained as
$$z_2 = \frac{1+z}{f_{z\delta}} - 1, \;\; \mbox{and}\;\;
z_3 = \frac{1+z}{f_{z\gamma}} - 1.$$ 
For this set of scale factors the directional Hubble parameters can be written 
as
\begin{equation}\label{dir_hub1_Z_caseII}
H_1= -\frac{\dot{z}}{1+z},\;\;H_2= -\,\frac{\dot{z}}{(1+z)f_{z\delta}},\;\;
H_3= -\,\frac{\dot{z}}{(1+z)f_{z\gamma}}.
\end{equation}
Thus the expression of average Hubble parameter takes the form:
\begin{equation}\label{avg_hub_caseII}
H=-\frac{\dot{z}}{3(1+z)}\Big[1+f_{z\delta}^{-1}+ f_{z\gamma}^{-1}\Big].
\end{equation}
In view of the above set of scale factors, Eq.\ (\ref{hubble_density_caseII}) 
can now be rewritten as
\begin{equation}\label{H_Z_caseII}
H= H_{0}\sqrt{E(z)},
\end{equation}
where
\begin{equation}
E(z) = \Omega_{m0}\big(f_{z\delta}f_{z\gamma}\big)^{-1}(1+z)^3 +\Omega_{r0}\big(f_{z\delta}f_{z\gamma}\big)^{-\frac{4}{3}}(1+z)^4 + \Omega_{\Lambda 0}+ 
\Omega_{\sigma 0}\big(f_{z\delta}f_{z\gamma}\big)^{-2} (1+z)^{6}.
\end{equation}
Correspondingly, the age of the universe can be written as
\begin{equation}\label{age_univ_caseII}
t_{age}=\frac{1}{H_0}\int_{0}^{\infty}\frac{dz}{(1+z)\sqrt{E(z)}}-\frac{1}{3H_0}\int_{0}^{\infty}\Big[\frac{\delta}{f_{z\delta}\sqrt{E(z)}}+ \frac{\gamma}{f_{z\gamma}\sqrt{E(z)}}\Big]dz.
\end{equation}
As before, other cosmological parameters in terms of redshift are as follows:\\

(i) Deceleration parameter:
\begin{equation}\label{dec_Z_caseII}
q(z)= \frac{1}{\sqrt{E(z)}}\Big[\frac{1}{2}\,\Omega_{m0}\big(f_{z\delta}f_{z\gamma}\big)^{-1}(1+z)^{3}+\Omega_{r0}\big(f_{z\delta}f_{z\gamma}\big)^{-\frac{4}{3}}(1+z)^{4}+\Omega_{\Lambda 0}+2\Omega_{\sigma 0}\big(f_{z\delta}f_{z\gamma}\big)^{-2}(1+z)^{6}\Big].
\end{equation}

(ii) Equation of State:
\begin{equation}\label{eos_Z_caseII}
\omega(z) = \frac{\frac{1}{3}\Omega_{r0}\big(f_{z\delta}f_{z\gamma}\big)^{-\frac{4}{3}}(1+z)^{4}-\Omega_{\Lambda 0}}{\Omega_{m0}\big(f_{z\delta}f_{z\gamma}\big)^{-1}(1+z)^{3}+\Omega_{r0}\big(f_{z\delta}f_{z\gamma}\big)^{-\frac{4}{3}}(1+z)^{4}+\Omega_{\Lambda 0}}.
\end{equation}

(iii) Ricci Scalar:
\begin{equation}\label{R_Z_caseII}
R(z)= 3H_0^2\Big[\Omega_{m0}\big(f_{z\delta}f_{z\gamma}\big)^{-1}(1+z)^{3}+4\,\Omega_{\Lambda 0}\Big].
\end{equation}

(iv) Luminosity distance:
\begin{equation}\label{lumin_Z_caseII}
d_L(z)=\frac{\big(f_{z\delta}f_{z\gamma}\big)^{-\frac{1}{3}}(1+z)}{H_0}\int_{0}^{\infty}\bigg[\frac{1}{\big(f_{z\delta}f_{z\gamma}\big)^{\frac{1}{3}}\sqrt{E(z)}}-\frac{\delta (1+z)}{3\,f_{z\delta}^{\frac{4}{3}}f_{z\gamma}^{\frac{1}{3}}\sqrt{E(z)}}-\frac{\gamma (1+z)}{3\,f_{z\delta}^{\frac{1}{3}}f_{z\gamma}^{\frac{4}{3}}\sqrt{E(z)}}\bigg]dz.
\end{equation}

(vi) Density parameter for anisotropy:
\begin{equation}\label{density_sigma_Z_caseII}
\Omega_{\sigma} = \frac{AB}{C E(z)},
\end{equation}
where $A = \Omega_{m0}\big(f_{z\delta}f_{z\gamma}\big)^{-1}(1+z)^{3}+\Omega_{r0}\big(f_{z\delta}f_{z\gamma}\big)^{-\frac{4}{3}} (1+z)^{4} + \Omega_{\Lambda 0}$,
$B = 1+f_{z\delta}^{-2}+f_{z\gamma}^{-2}-\Big[f_{z\delta}^{-1}+f_{z\gamma}^{-1}+\big(f_{z\delta}f_{z\gamma}\big)^{-1}\Big]$ and 
$ C = f_{z\delta}^{-1}+f_{z\gamma}^{-1}+\big(f_{z\delta}f_{z\gamma}\big)^{-1}$.
\\

(vii) Density parameter for matter:
\begin{equation}\label{dens_mat_II}
\Omega_m(z) = \frac{\Omega_{m0}\big(f_{z\delta}f_{z\gamma}\big)^{-1}(1+z)^{3}}{E(z)}.
\end{equation}

{As in the previous case, we have obtained these forms of expressions of 
various cosmological parameters by considering the scale factors as 
given in Eq.~{\eqref{scale1_caseII_ad}}. The detailed graphical analysis 
of these expressions are presented in the next section.}

\section{Graphical analysis and results}\label{4}

In this section we focus on the graphical analysis of various cosmological 
parameters discussed above and try to give possible explanations of the 
results obtained from the emerging graphs. Here, in the numerical calculations, we have taken 
the Planck 2018 results on cosmological parameters \citep{Planck_2018}.

\subsubsection{\textbf{Hubble parameter}}
\begin{figure}[!htb]
\centerline{
  \includegraphics[scale = 0.35]{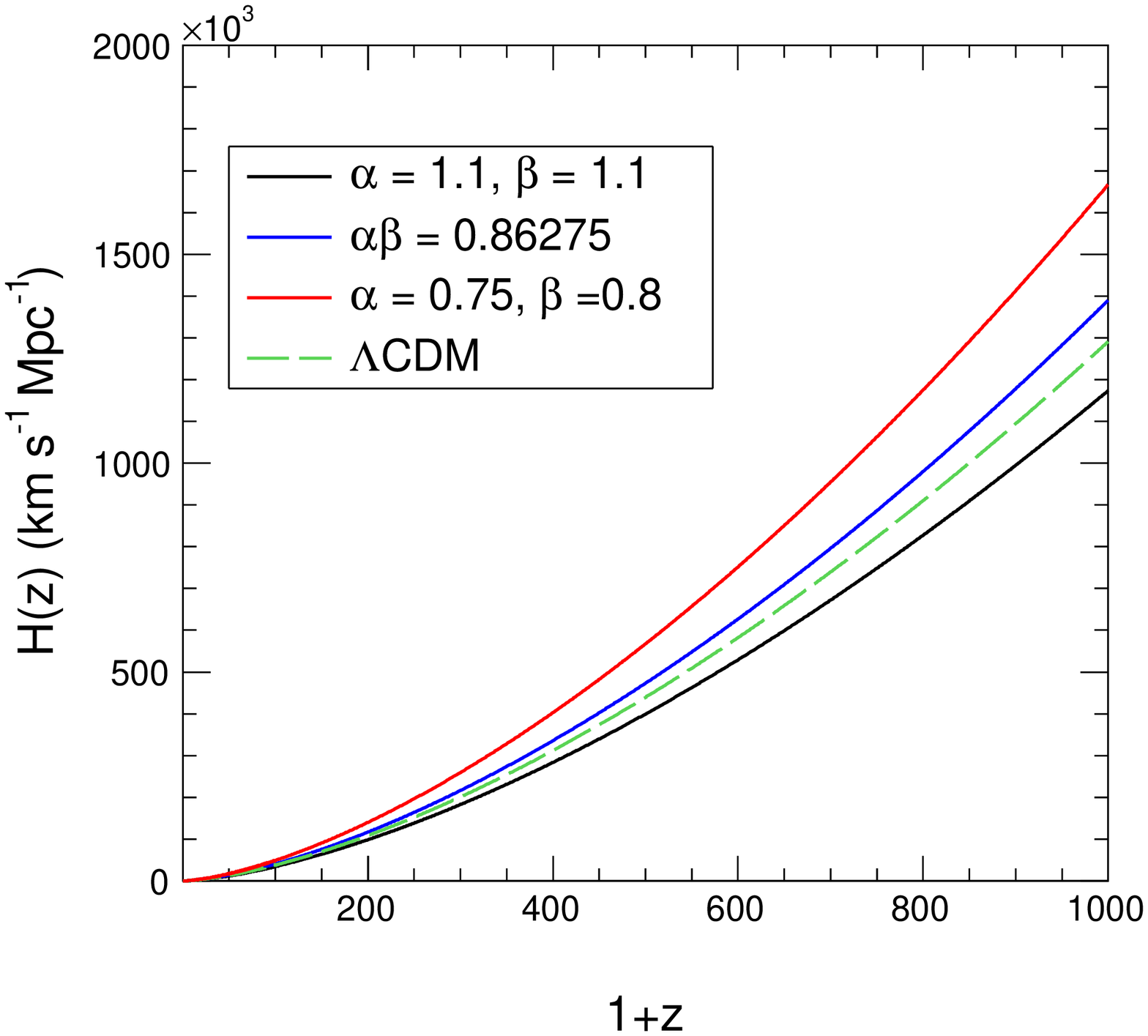}\hspace{1cm}
  \includegraphics[scale = 0.35]{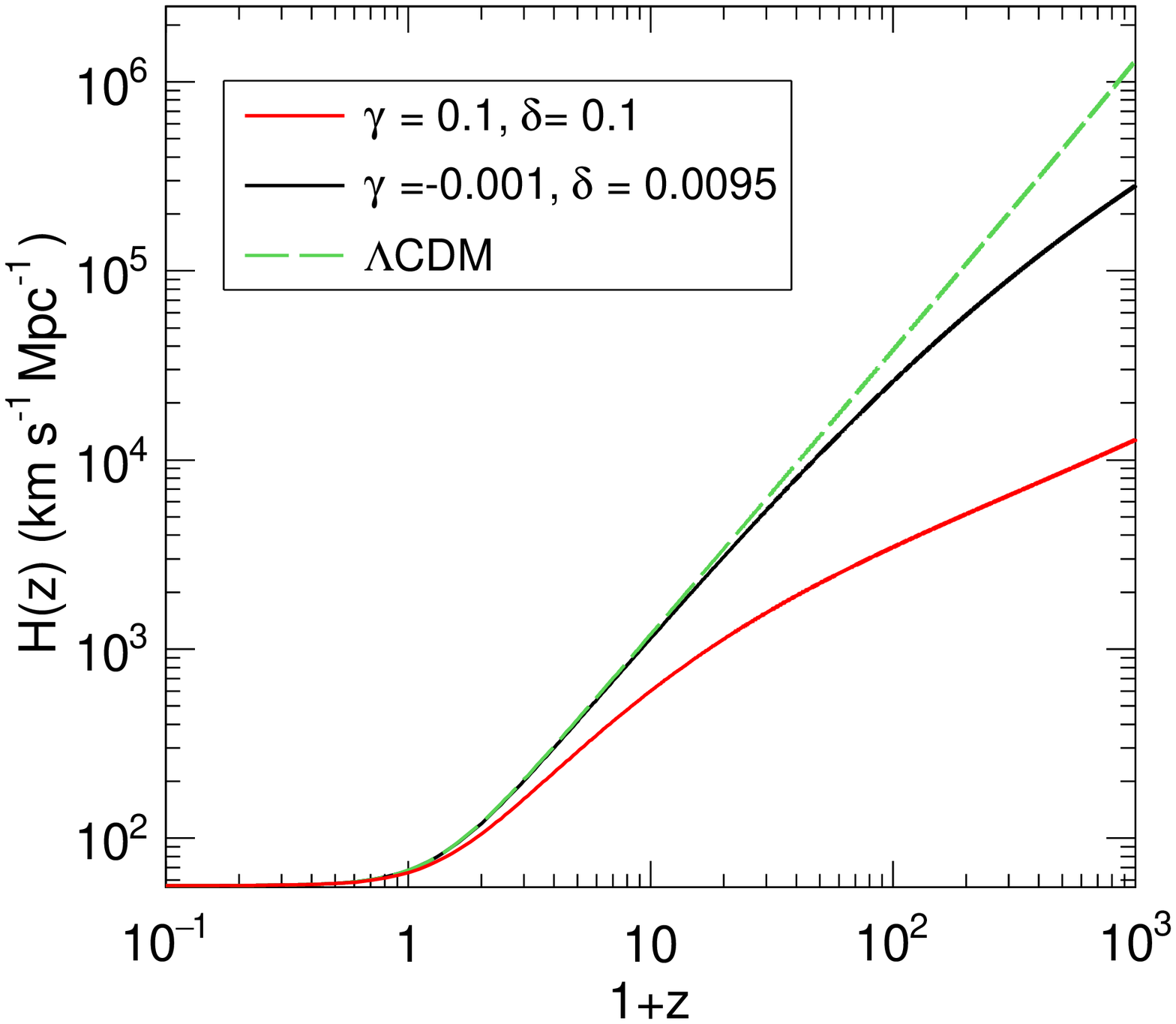}} 
\caption{Behaviour of Hubble parameter $H(z)$ with respect to $1+z$ for the
case I (left) and case II (right) with different associated model parameters.}
\label{fig1a}
\end{figure}

To understand the behaviour of Hubble parameter $H(z)$ with respect to $z$, we 
first plot the values of $H(z)$ with $1+z$ over a large range of its values for three different sets of values of $\alpha$ and $\beta$ for the case I and 
two sets of values of $\gamma$ and $\delta$ for the case II as shown in the 
Fig.\ \ref{fig1a} by using Eqs.\ \eqref{H_Z_caseI} and \eqref{H_Z_caseII} 
respectively. As mentioned earlier, it is seen from the figure that both the
cases deviate significantly from the isotropic universe ($\Lambda$CDM model) in
the remote past, whereas they agree with the isotropic universe at present and
at distant future epochs. However, there is a difference of variation 
patterns of $H(z)$ in between these two cases in the past. In the case I, the 
values of $H(z)$ can be greater or smaller than its corresponding values in 
the $\Lambda$CDM model in the past depending on the values of parameters 
$\alpha$ and $\beta$. On the other hand in the case II, the values of $H(z)$ 
never greater than its corresponding values in the $\Lambda$CDM model in the 
past for any set of values of parameters $\gamma$ and $\delta$. Also we have 
noticed that, more the values of $\alpha$ and $\beta$ moves towards one, higher
the tendency of the Hubble parameter plot in case I to move towards the 
$\Lambda$CDM model. Whereas, smaller values of $\delta$ and $\gamma$ show 
excellent agreement with the $\Lambda$CDM plot.    

\begin{figure}[!htb]
\centerline{
  \includegraphics[scale = 0.35]{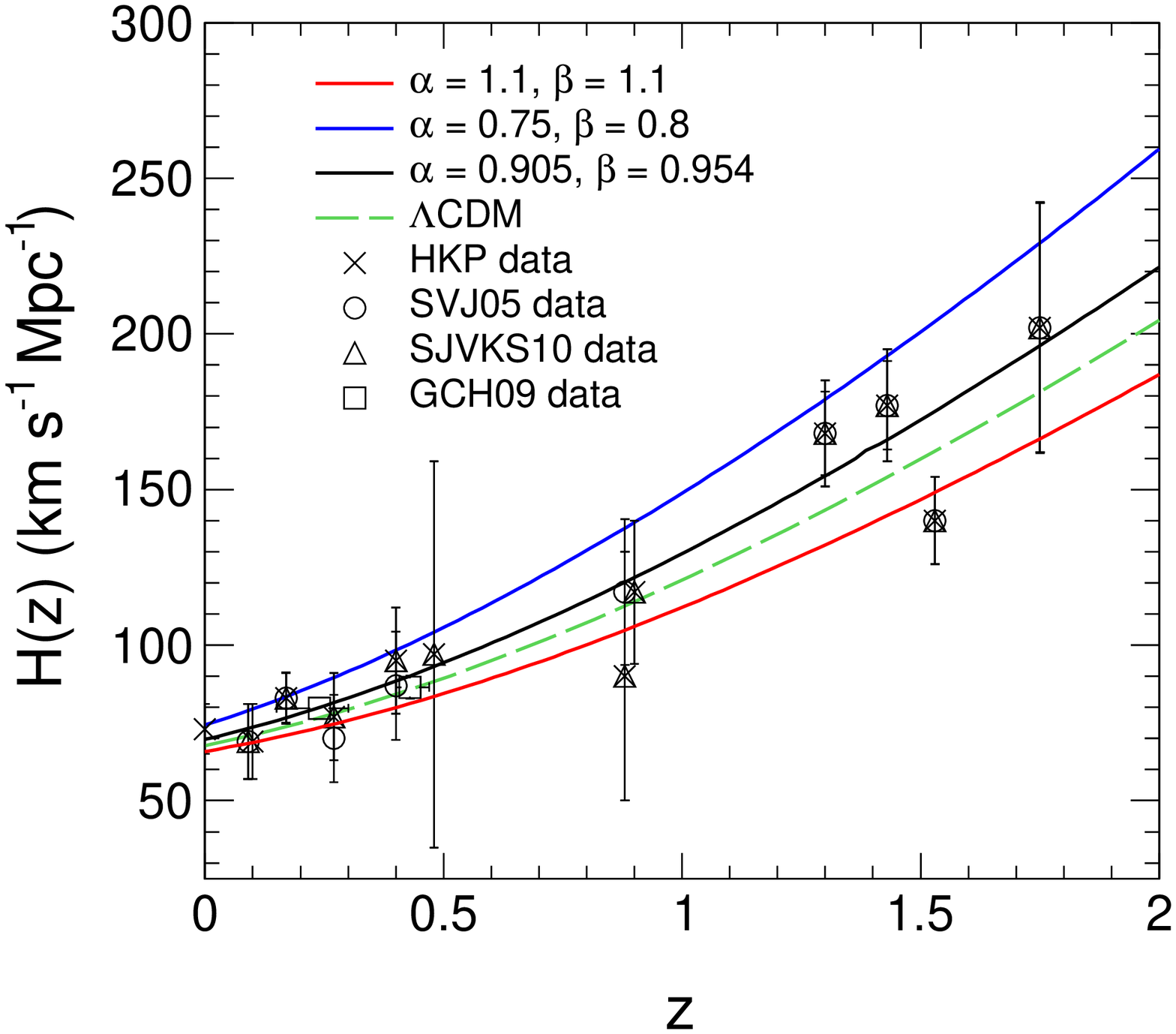}\hspace{1cm}
  \includegraphics[scale = 0.35]{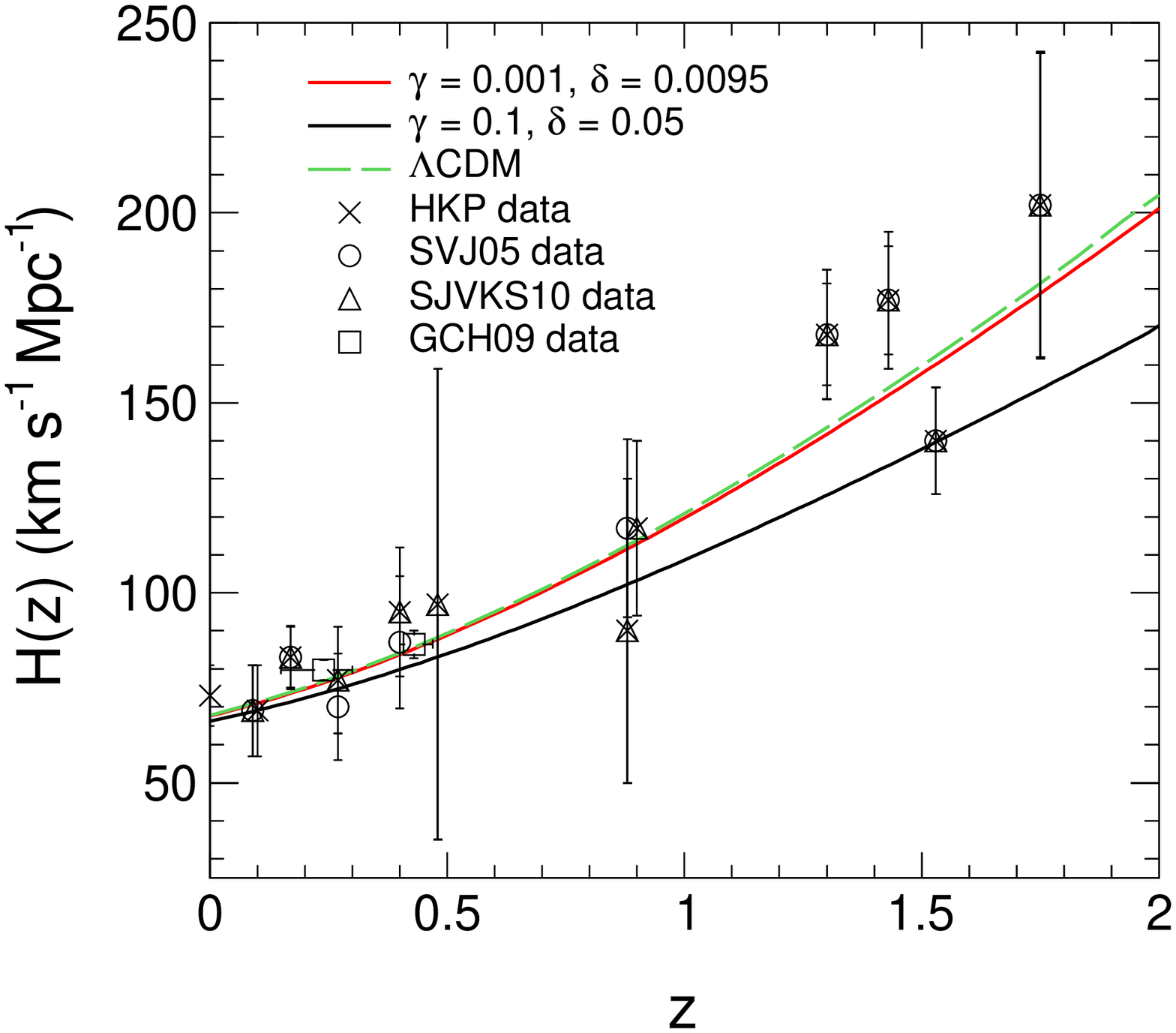}} 
\caption{Behaviour of Hubble parameter $H(z)$ for the small $z$ values with
different values of model parameters in conformity with four sets of observed 
data within the available ranges of errors for the case I (left) and case II 
(right).}
\label{fig1b}
\end{figure}

Moreover, to choose a reliable range of values of model parameters, i.e.\ 
$\alpha, \beta$ for the case I and $\gamma, \delta$ for the case II, we have 
used four sets of available $H(z)$ data, viz., HKP and SVJ05 data 
\cite{Simon_2005}, SJVKS10 data \cite{Stern_2010} and GCH09 data 
\cite{Gaztanaga_2009} in the $H(z)$ versus $z$ plots of Fig.\ \ref{fig1b} to 
constrained the model parameters within the observable range of $z$ values. 
Since the SVJ05 data set has been replaced already by SJVKS10 data, we have 
taken this data set only as a reference \cite{Ma_2011}. From this figure, we 
have found a good range of values of the product of the two free parameters
(i.e.\ $\alpha\beta$) from 0.6 to 1.21 for the case I. For different sets of 
values of $\alpha$ and $\beta$, whose products lie in between the above 
mentioned range of values will give us consistent results with the 
observational data. In this paper, we have taken three different sets of values 
of $\alpha$ and $\beta$: (0.75, 0.8), (0.905, 0.954) and (1.1, 1.1) as shown 
in the left plot of Fig.\ \ref{fig1b}. For the case II, we have found a best 
possible range of values of $\gamma$ and $\delta$ as (0.001, 0.0095) to 
(0.1, 0.05), within which any set of values of $\gamma$ and $\delta$ gives 
results that fit to data within range of error bars. It is also seen that the 
set $\gamma = 0.001$ and $\delta = 0.0095$ produces the results that almost 
agree with the results of the $\Lambda$CDM model upto the values of $z\sim 1.2$.
\begin{figure}[!htb]
 \centerline{
   \includegraphics[scale = 0.35]{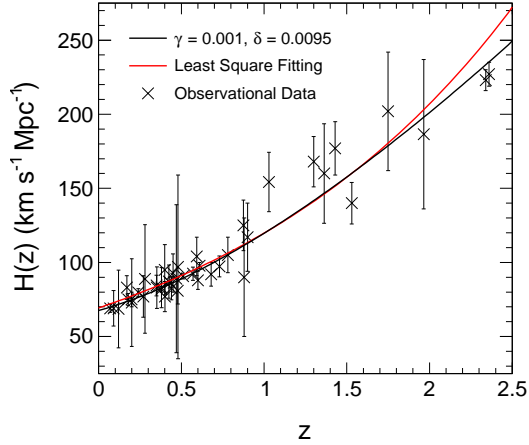}}
\caption{Least square fitting curve and the best fitted curve of $H(z)$
obtained from Eq.\ \eqref{H_Z_caseII} for the values of the parameters
$\delta$ and $\gamma$ as 0.001 and 0.0095 respectively to the observed Hubble
parameter data set shown in Table \ref{tableI}.}
\label{fig2}
\end{figure}

Further, to get the precise values of above free parameters for both the cases,
which should be consistent with the $\Lambda$CDM model at least around the
present epoch, we have considered the compilation of 43 observational Hubble
parameter data against 43 different values of redshift \cite{Akarsu_2019, 
Gogoi_2021} as shown in Table \ref{tableI}. We have plotted the best fitted
curve to this set of Hubble parameter data against the redshift by using a
non-linear curve fitting (least square fitting) technique as shown in
Fig.\ \ref{fig2}. From the curve we have calculated the value of
$\alpha\beta = 0.86275$ for the case I at $z=0$. It is important to mention
that we may be able to consider a set of values of $\alpha$ and $\beta$, which
satisfy the condition that $\alpha\beta = 0.86275$. Thus, we can use different
sets of $\alpha$ and $\beta$ with this condition. Here, we have not consider
any special set of values of $\alpha$ and $\beta$, but just considered the
condition $\alpha\beta = 0.86275$ in the rest of the paper. However, it is not
easy to get the values of the parameters $\delta$ and $\gamma$ for the case II
using this method. So, for the case II we have used a suitable set of values of
$\delta$ and $\gamma$, for which the expression of $H(z)$ for this case
(Eq.\ \eqref{H_Z_caseII}) gives a curve very close to the least square fitting
curve to the observational data as shown in Fig.\ \ref{fig2}. This set of
values of $\delta$ and $\gamma$ are found to be 0.001 and 0.0095 respectively,
which is the same as the lower limit values of these parameters found from the
analysis with Fig.\ \ref{fig1b}. Thus the graphical plots of $H(z)$ for both the
cases obtained from these derived free parameters values as mentioned above
should be consistent with the $\Lambda$CDM plot shown in Fig.\ \ref{fig1b},
especially for the range of small $z$.
\begin{center}
\begin{table}[!hbt]
\caption{Presently available observational Hubble parameter ($H^{obs}(z)$) 
data set. $H^{obs}(z)$ is in unit of km/s/Mpc.}
\vspace{2mm}
\begin{tabular}{ccc|ccc}
\hline 
\rule[-1ex]{0pt}{2.5ex} \hspace{0.5cm} $z$ \hspace{0.5cm}  & \hspace{0.5cm} $ H^{obs}(z)$ \hspace{0.5cm} & \hspace{0.5cm} Reference \hspace{0.5cm} &
\hspace{0.5cm} $z$ \hspace{0.5cm}  & \hspace{0.5cm} $ H^{obs}(z)$ \hspace{0.5cm} & \hspace{0.5cm} Reference \hspace{0.5cm} \\ 
\hline
\rule[-1ex]{0pt}{2.5ex} 0.0708 & 69.0 $\pm$ 19.68 & \cite{Zhang_2014} &
0.48 & 97.0 $\pm$ 62.0 & \cite{Ratsimbazafy_2017}\\
\rule[-1ex]{0pt}{2.5ex} 0.09 & 69.0 $\pm$ 12.0 & \cite{Simon_2005} &
0.51 & 90.8 $\pm$ 1.9 & \cite{Alam_2017}\\
\rule[-1ex]{0pt}{2.5ex} 0.12 & 68.6 $\pm$ 26.2 & \cite{Zhang_2014} &
0.57 & 92.4 $\pm$ 4.5 & \cite{Samushia_2013} \\
\rule[-1ex]{0pt}{2.5ex} 0.17 & 83.0 $\pm$ 8.0 & \cite{Simon_2005} &
0.593 & 104.0 $\pm$ 13.0 & \cite{Moresco_2012}\\
\rule[-1ex]{0pt}{2.5ex} 0.179 & 75.0 $\pm$ 4.0 & \cite{Moresco_2012} &
0.60 & 87.9 $\pm$ 6.1 & \cite{Blake_2012}\\
\rule[-1ex]{0pt}{2.5ex} 0.199 & 75.0 $\pm$ 5.0 & \cite{Moresco_2012} &
0.61 & 97.8 $\pm$ 2.1 & \cite{Alam_2017}\\
\rule[-1ex]{0pt}{2.5ex} 0.20 & 72.9 $\pm$ 29.6 & \cite{Zhang_2014} &
0.68 & 92.0 $\pm$ 8.0 & \cite{Moresco_2012}\\
\rule[-1ex]{0pt}{2.5ex} 0.24 & 79.69 $\pm$ 2.65 & \cite{Gaztanaga_2009} &
0.73 & 97.3 $\pm$ 7.0 & \cite{Blake_2012}\\
\rule[-1ex]{0pt}{2.5ex} 0.27 & 77.0 $\pm$ 14.0 & \cite{Simon_2005} &
0.781 & 105.0 $\pm$ 12.0 & \cite{Moresco_2012}\\
\rule[-1ex]{0pt}{2.5ex} 0.28 & 88.8 $\pm$ 36.6 & \cite{Zhang_2014} &
0.875 & 125.0 $\pm$ 17.0 & \cite{Moresco_2012}\\
\rule[-1ex]{0pt}{2.5ex} 0.35 & 84.4 $\pm$ 7.0  & \cite{Xu_2013} &
0.88 & 90.0 $\pm$ 40.0 & \cite{Ratsimbazafy_2017}\\
\rule[-1ex]{0pt}{2.5ex} 0.352 & 83.0 $\pm$ 14.0 & \cite{Moresco_2012}&
0.90 & 117.0 $\pm$ 23.0 & \cite{Simon_2005}\\
\rule[-1ex]{0pt}{2.5ex} 0.38 & 81.9 $\pm$ 1.9 & \cite{Alam_2017} &
1.037 & 154.0 $\pm$ 20.0 & \cite{Moresco_2012}\\
\rule[-1ex]{0pt}{2.5ex} 0.3802 & 83.0 $\pm$ 13.5 & \cite{Moresco_2016} &
1.30 & 168.0 $\pm$ 17.0 & \cite{Simon_2005}\\
\rule[-1ex]{0pt}{2.5ex} 0.40 & 95.0 $\pm$ 17.0 & \cite{Simon_2005} &
1.363 & 160.0 $\pm$ 33.6 & \cite{Moresco_2015}\\
\rule[-1ex]{0pt}{2.5ex} 0.4004 & 77.0 $\pm$ 10.2 & \cite{Moresco_2016} &
1.43 & 177.0 $\pm$ 18.0 & \cite{Simon_2005}\\
\rule[-1ex]{0pt}{2.5ex} 0.4247 & 87.1 $\pm$ 11.2 & \cite{Moresco_2016} &
1.53 & 140.0 $\pm$ 14.0 & \cite{Simon_2005}\\
\rule[-1ex]{0pt}{2.5ex} 0.43 & 86.45 $\pm$ 3.68 & \cite{Gaztanaga_2009} &
1.75 & 202.0 $\pm$ 40.0 & \cite{Simon_2005}\\
\rule[-1ex]{0pt}{2.5ex} 0.44 & 82.6 $\pm$ 7.8 & \cite{Blake_2012} &
1.965 & 186.5 $\pm$ 50.4 & \cite{Moresco_2015}\\
\rule[-1ex]{0pt}{2.5ex} 0.4497 & 92.8 $\pm$ 12.9 & \cite{Moresco_2016} & 
2.34 & 223.0 $\pm$ 7.0 & \cite{Delubac_2015}\\
\rule[-1ex]{0pt}{2.5ex} 0.47 & 89.0 $\pm$ 50.0 & \cite{Ratsimbazafy_2017} &
2.36 & 227.0 $\pm$ 8.0 & \cite{Ribera_2014}\\
\rule[-1ex]{0pt}{2.5ex} 0.4783 & 80.9 $\pm$ 9.0 & \cite{Moresco_2016} &&&\\
\hline
\end{tabular}
\label{tableI}
\end{table}
\end{center}

\subsubsection{\textbf{Ricci scalar}}
\begin{figure}[!htb]
\centerline{
  \includegraphics[scale = 0.35]{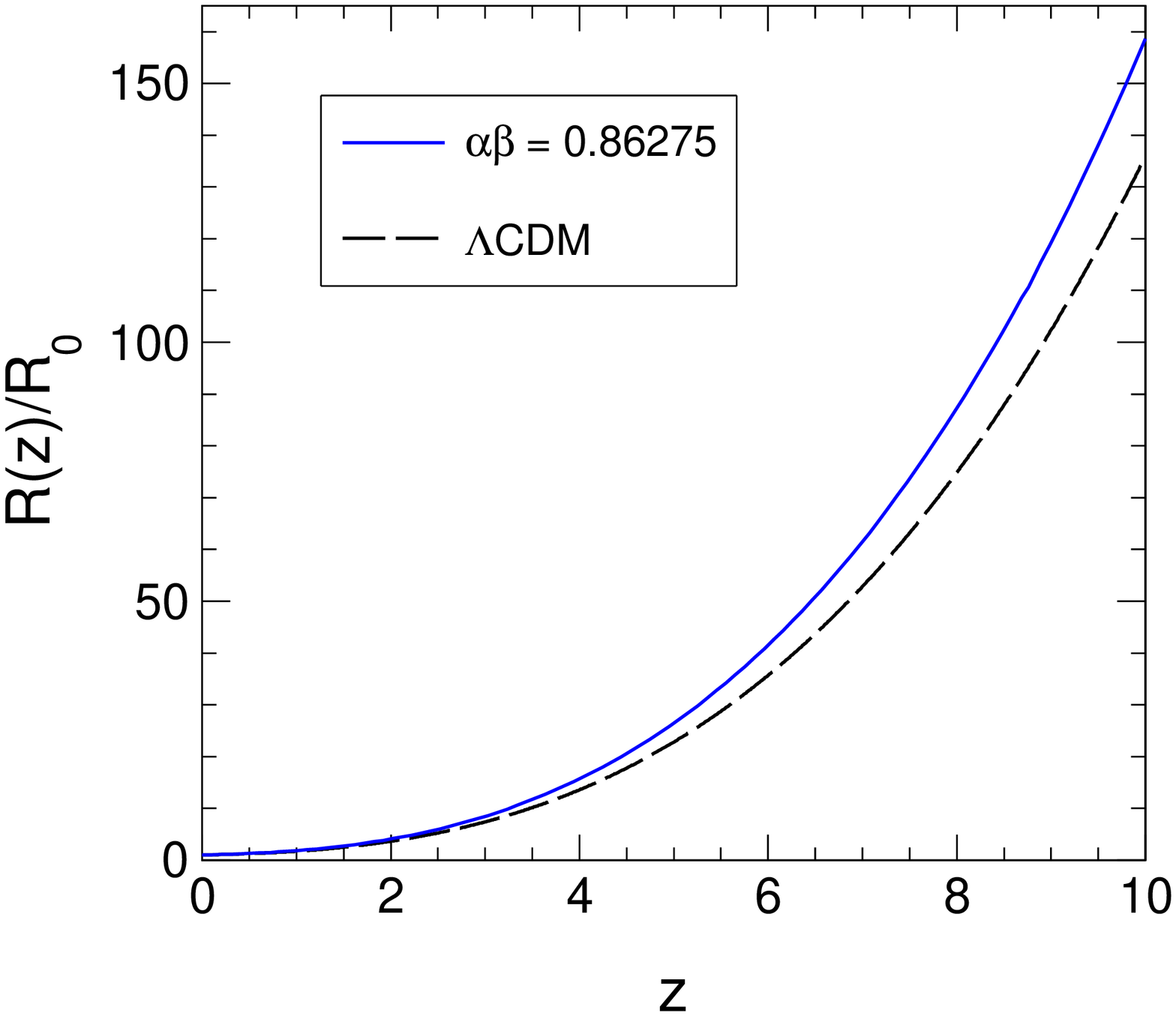}\hspace{1cm} 
  \includegraphics[scale = 0.35]{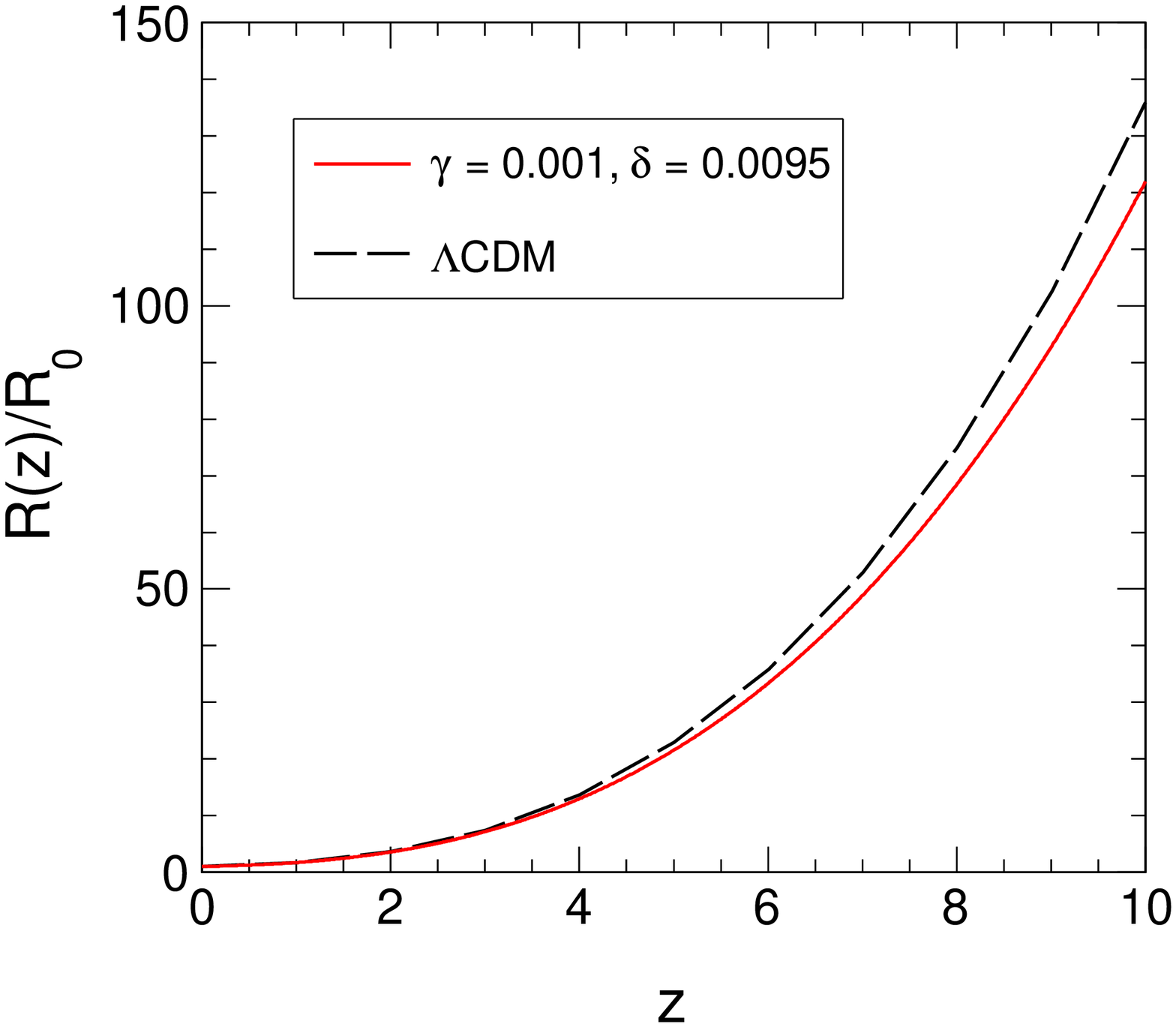}} 
\caption{Ricci scalar versus redshift plots for the case I (left) and case II
(right) of the proposed anisotropic model of the universe with the constrained 
values of free model parameters along with the plot for the $\Lambda$CDM model.}
\label{fig3}
\end{figure}

Ricci scalar is an important parameter to specify the geometric property of 
spacetime of the universe. {Some mathematical and graphical analysis of 
the evolution of Ricci scalar for various cosmological models in modified 
cosmology are also found in Ref.} \cite{Gogoi_2021}. So, to understand the 
evolution of Ricci scalar in 
our anisotropic model of the universe, we have plotted Ricci scalar $R(z)$ 
against the redshift ($z$) for the case I and case II by using Eqs.\ 
(\ref{R_Z_caseI}) and (\ref{R_Z_caseII}) respectively as shown in 
Fig.\ \ref{fig3}. In this plot we have used the observationally constrained 
value of $\alpha\beta = 0.86275$ for the case I, and $\gamma = 0.001$ and  
$\delta = 0.0095$ for the case II as mentioned in the previous subsection. 
Here, we have also shown the plot of this cosmological parameter for the 
$\Lambda$CDM model. Figure shows that both the cases of our model are 
consistent with the $\Lambda$CDM model from the present epoch to the near past.
However, the deviation of both the cases increases with higher curvature of 
spacetime for the case I and with lower curvature for the case II, from the 
$\Lambda$CDM model as we move towards the distant past.  

\subsubsection{\textbf{Deceleration Parameter and Equation of State}}
\begin{figure}[!htb]
\centerline{
  \includegraphics[scale = 0.35]{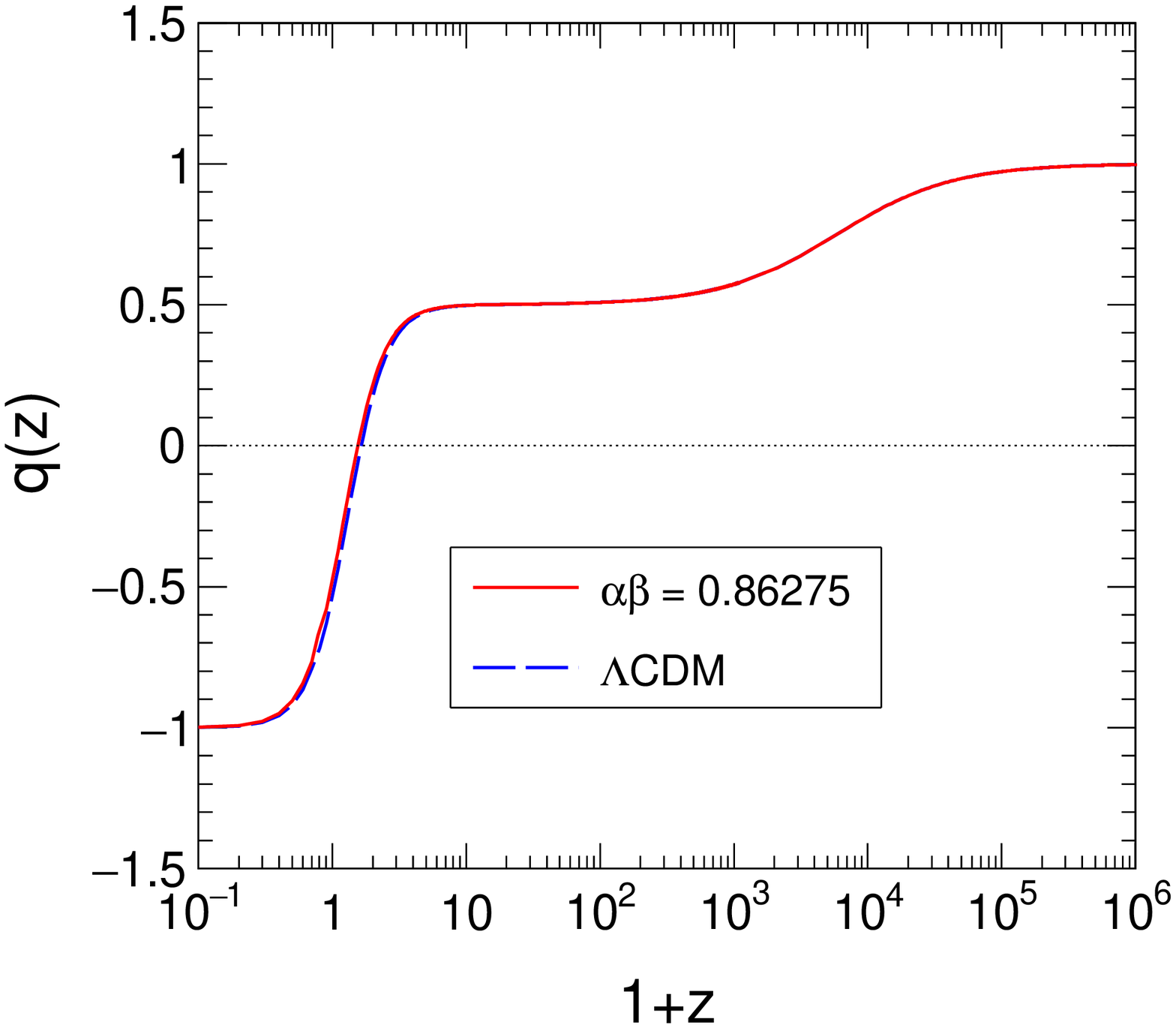}\hspace{1cm} 
  \includegraphics[scale = 0.35]{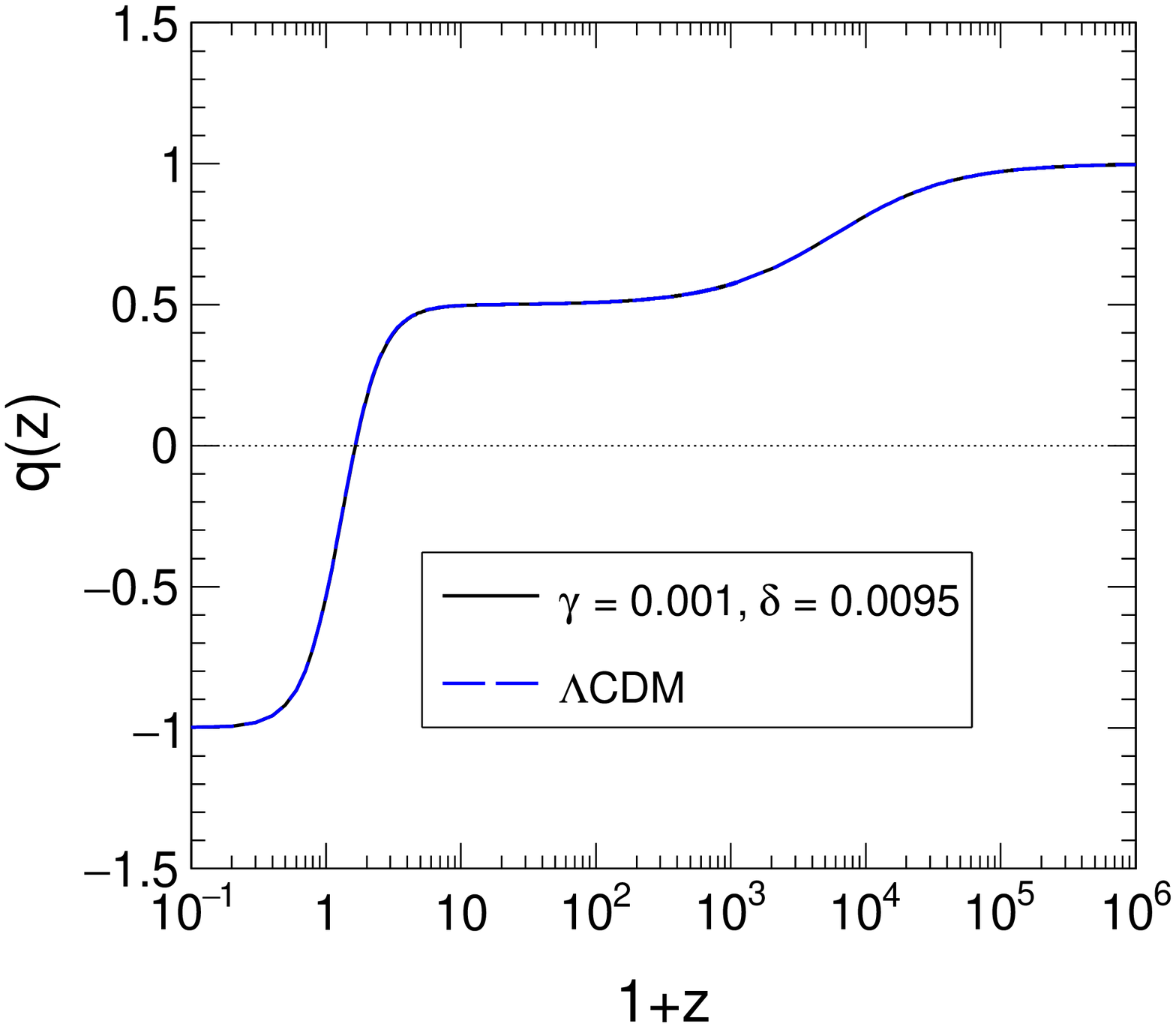}} 
\caption{Deceleration parameter versus redshift plots for the case I (left) and 
case II (right) of anisotropic model of the universe with the constrained 
values free model parameters along with the plot for the $\Lambda$CDM model.}
\label{fig4}
\end{figure}

As the deceleration parameter $q(z)$ is an important parameter to understand 
the pattern of evolution of the universe, we have plotted it with respect to 
$1+z$ for the both the cases of our model for the constrained values of 
the free parameters as mentioned above by using Eqs.\ (\ref{q_Z_caseI}) and 
(\ref{dec_Z_caseII}) along with the plot for the $\Lambda$CDM model as shown in 
Fig.\ \ref{fig4}. It is found that for both the cases, at $z$ = 0, $q(0)$ 
tends to $-0.55$, which is a good agreement with the $\Lambda$CDM model's 
prediction for the current value of $q(z)$. In reality for any value of $z$, 
$q(z)$ for both the cases shows a good agreement with the $\Lambda$CDM model, 
especially the case II has the excellent agreement. This implies that the 
pattern of evolution of our anisotropic model of universe is same as that of 
the isotropic model.    

\begin{figure}[!htb]
\centerline{
  \includegraphics[scale = 0.35]{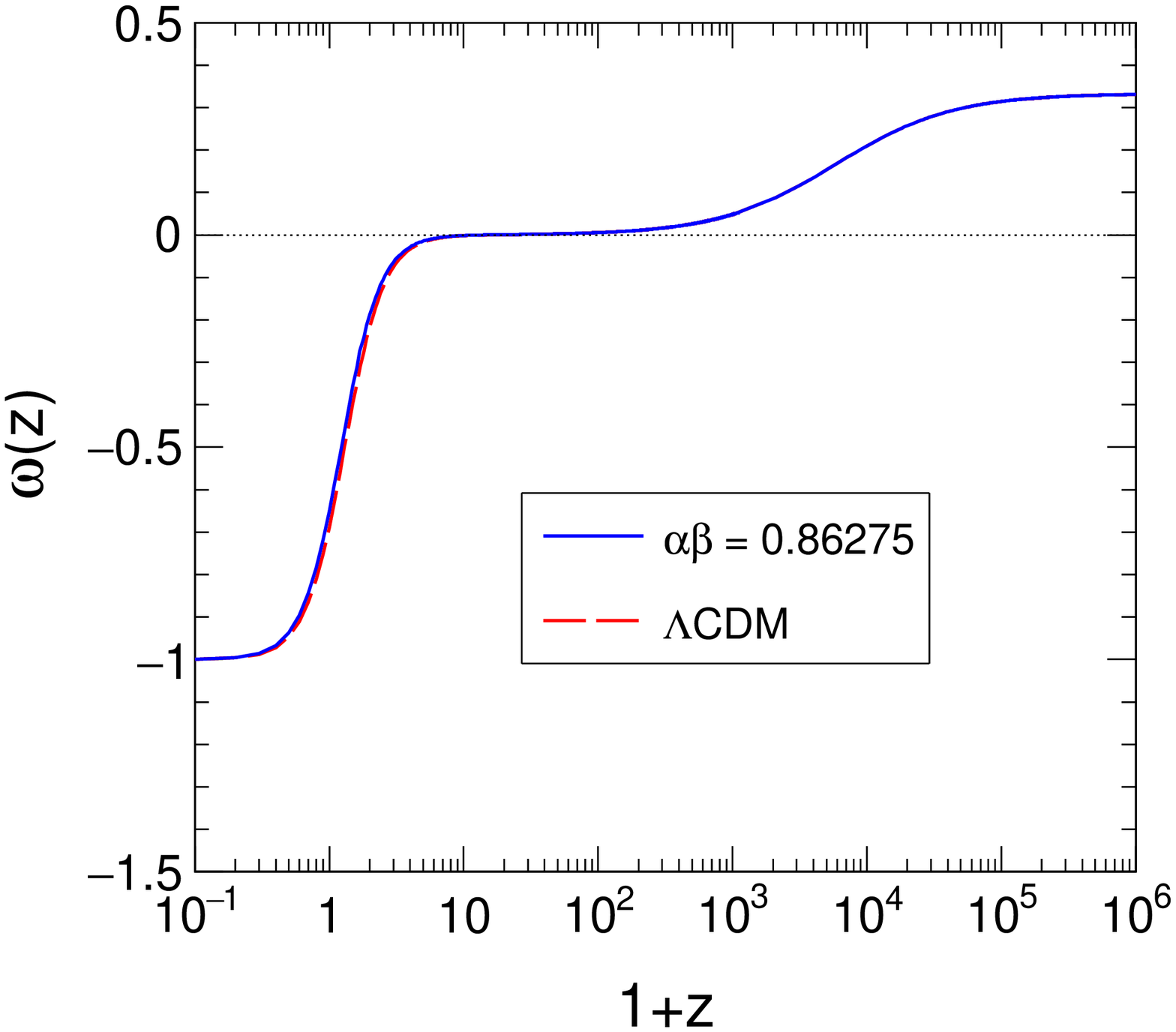}\hspace{1cm} 
  \includegraphics[scale = 0.35]{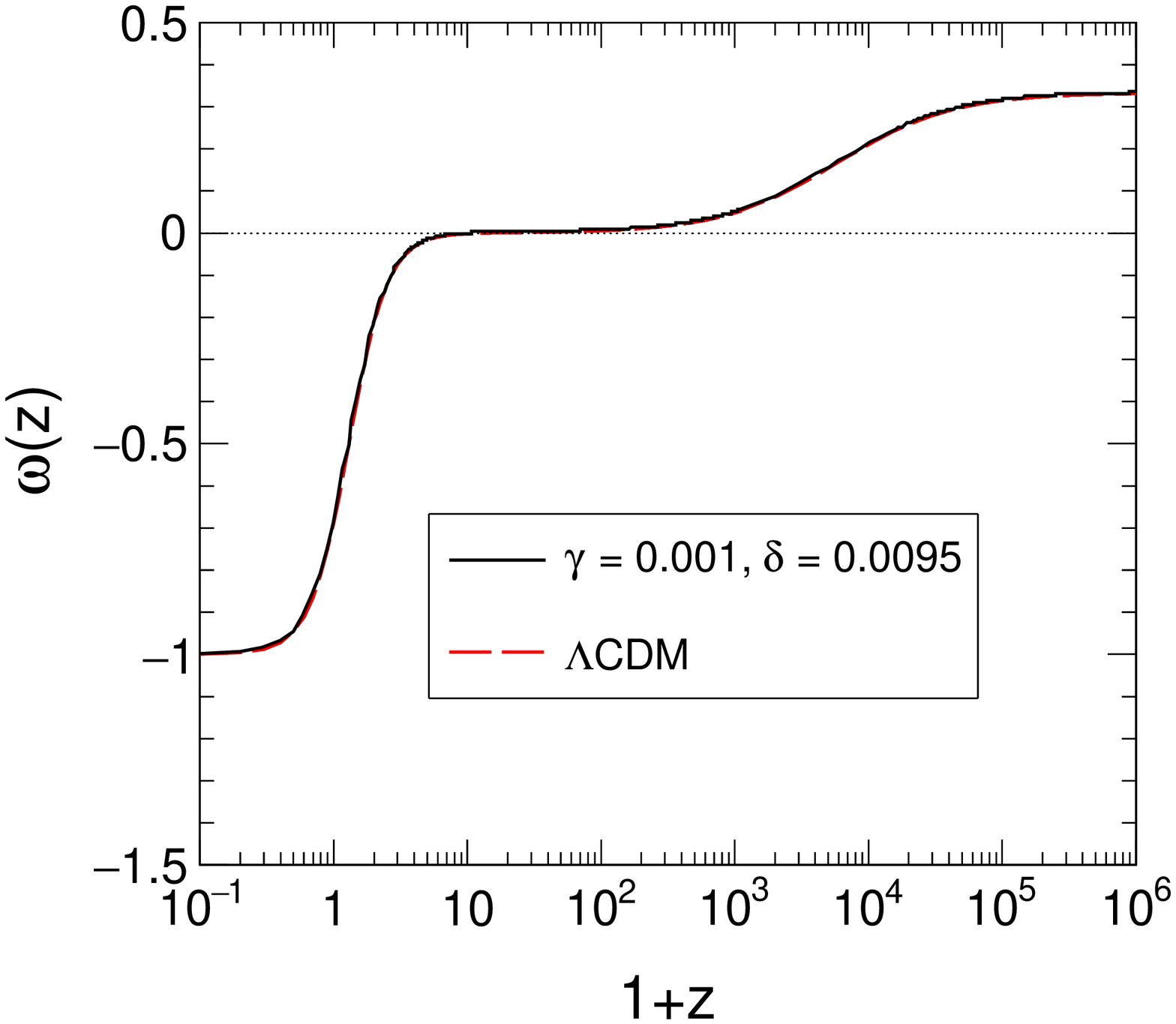}} 
\caption{Equation of State parameter with respect to redshift for the 
case I (left) and case II (right) of anisotropic model of the universe with 
the constrained values free model parameters along with the plot for the 
$\Lambda$CDM model.}
\label{fig5}
\end{figure}

Similarly, the Equation of State (EoS) is an important cosmological parameter 
to understand the composition of the universe as well as its characteristics. 
We have plotted the EoS ($\omega$) against $1+z$ for both the cases by 
using Eq.\ (\ref{eos_Z_caseI}) and (\ref{eos_Z_caseII}) with the constrained
values of free model parameters as shown in Fig.\ \ref{fig5}. We found that 
at $z=0$, EoS $\omega$ is close to $-0.7$ for both the cases, which is also 
consistent with the $\Lambda$CDM model and agrees with the current era of the 
universe as the dark energy era. The curves show good agreement with the 
$\Lambda$CDM model (in fact it is excellent for the case II) and hence support 
the three phases of the universe, viz., the radiation dominated phase 
($\omega= 1/3$), the matter dominated phase ($\omega =0$) and the dark energy 
phase ($\omega=-1$). Thus this anisotropic model of the universe has the same
evolution characteristics as the isotropic model of the universe. 

\subsubsection{\textbf{Distance modulus}}
\begin{figure}[!htb]
\centerline{
  \includegraphics[scale = 0.35]{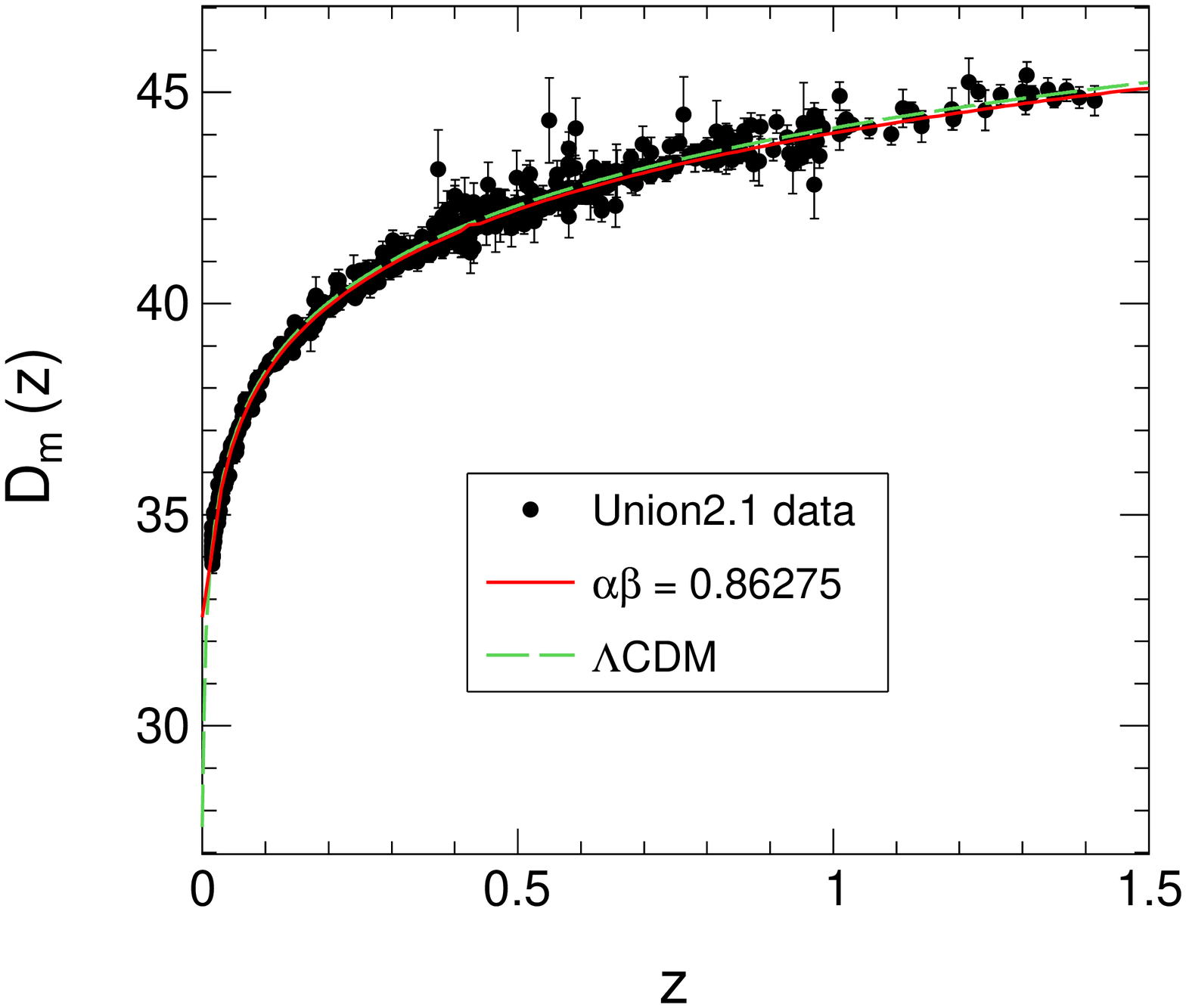}\hspace{1cm} 
  \includegraphics[scale = 0.35]{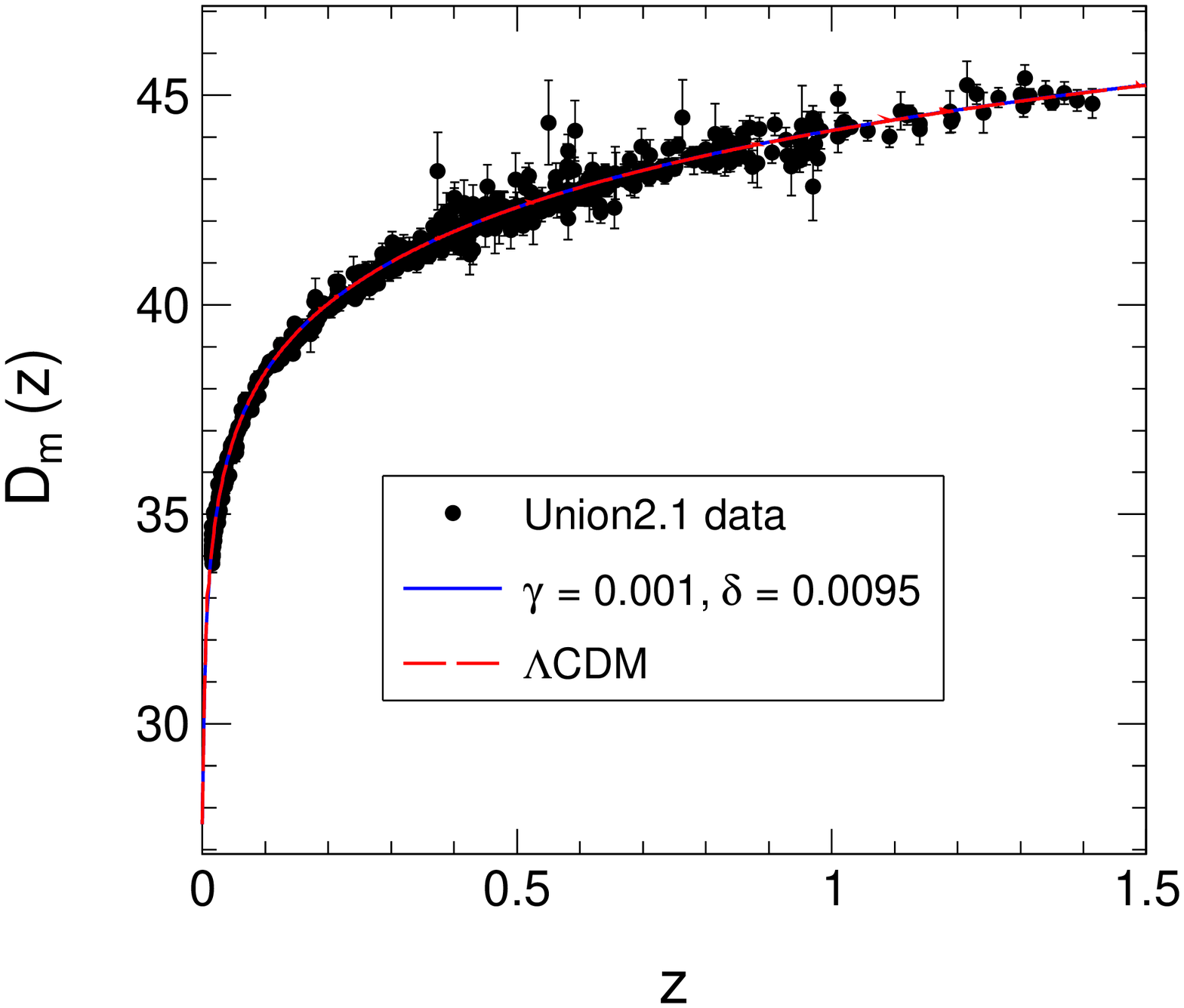}}
\caption{Distance modulus versus redshift plots for the case I (left) and 
case II (right) of anisotropic model of the universe with the constrained 
values free model parameters in comparison with Union2.1 data along with the 
plot for the $\Lambda$CDM model.}
\label{fig6}
\end{figure}

The distance modulus is a reliable observational parameter to understand the 
cosmic evolution with redshift. Here, we have plotted the distance modulus 
($D_m$) against redshift for both the cases with the free model parameters as
mention above by using Eqs.\ (\ref{Lumin_Z_caseI}) and (\ref{lumin_Z_caseII}) 
in Eq.\ (\ref{dist_mod}) along with the Union2.1 observational data 
\cite{Suzuki_2012} and $D_m$ for the $\Lambda$CDM model as shown in Fig.\ 
\ref{fig6}. It is clear from the figure that for both the cases, for the 
constrained free parameter values, the distance modulus plot shows excellent 
agreement with observational data along with the $\Lambda$CDM plot. This 
implies that the anisotropic model of the universe agrees very well with 
the observed data of the distance modulus of galaxies. 

\subsubsection{\textbf{Anisotropy density parameter}}
\begin{figure}[!htb]
\centerline{
  \includegraphics[scale=0.35]{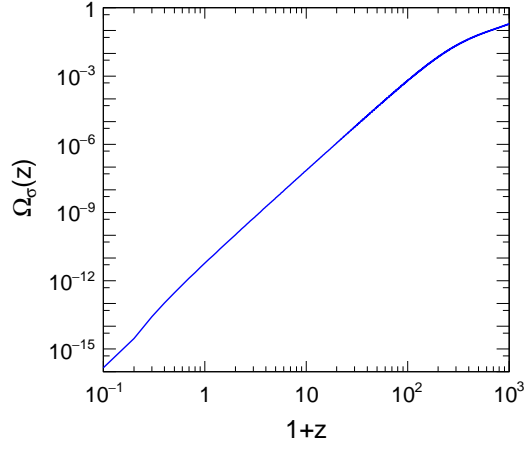}} 
\caption{Anisotropy density parameter versus redshift plot for the case II
with the constrained values of free model parameters $\gamma = 0.001$ and 
$\delta = 0.0095$.}
\label{fig7}
\end{figure}

The anisotropy density parameter is an important parameter to understand 
the possible anisotropy that exists in the universe. For the case I it's value 
is zero, but for the case II it has a definite expression as mentioned in
Eq.\ (\ref{density_sigma_Z_caseII}). Fig.\ \ref{fig7} shows the plot of this 
$\Omega_\sigma$ against $1+z$ for the constrained values of $\delta$ and 
$\gamma$. This figure suggests that a considerable amount of 
anisotropy exists for a much higher value of redshift, which indicates that the 
universe had prominent anisotropic character in its early stages and which is 
now almost reduced to  zero. {The existence of anisotropy after inflation 
period can be explained with the idea like breaking the Lorentz invariance by 
introducing a condensation of vector field} \cite{Kanno_2008}{, slow roll 
phase of vector fields like as inflation field in chaotic inflationary 
scenario,} \cite{Golovnev_2008,Kanno_2008,Watanabe_2009} {or considering 
anisotropic inflation model with vector impurity}\cite{Kanno_2008} {etc. 
However, this is a broad area of study and hence to be considered latter.} In 
Ref.\ \cite{Akarsu_2019}, the upper limits 
of the present value of the anisotropy density parameter ($\Omega_{\sigma 0}$) 
has been calculated by using the BAO and CMB data for the matter dominated 
recombination era as $\Omega_{\sigma 0}\le 10^{-15}$, and by demanding that
the standard big bang nucleosynthesis (BBN) is not significantly affected by the
expansion anisotropy as $\Omega_{\sigma 0}\le 10^{-23}$. From these two upper 
limits of $\Omega_{\sigma 0}$ we have calculated the corresponding values of 
the free parameters $\delta$ and $\gamma$ of the case II, and found that their
values may lie in between $10^{-3}$ and $10^{-8}$. It should be noted that this
calculated upper limit of values of $\delta$ and $\gamma$ is in agreement with 
their constrained values obtained from the Hubble parameter data as mentioned 
above.     

\subsubsection{\textbf{Relative shear anisotropy parameter}} 
\begin{figure}[!htb]
\centerline{
  \includegraphics[scale = 0.35]{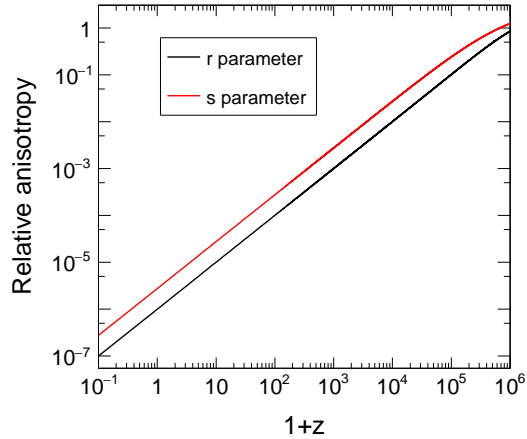}} 
\caption{Relative shear anisotropy versus redshift plot for the case II.}
\label{fig8}
\end{figure}

The relative shear anisotropy parameters $r$ and $s$ can be defined as \cite{Barrow_1997}
$$r = \frac{H_1-H_2}{H},\;\; s = \frac{H_1-H_3}{H},$$


where as mentioned above $H_1, H_2, H_3$ are the directional Hubble parameters 
and $H$ is the average Hubble parameter. The parameters $r$ and $s$ provide 
the information about if the anisotropy exists in different directions and 
their plot against redshift gives the picture about the evolution of 
directional anisotropy. Fig.\ \ref{fig8} shows the plots of both the 
parameters for case II. The figure shows that there was a considerable amount 
of relative shear anisotropy present for the higher value of redshift $(z)$. 
Thus the model predicts the existence of higher value of relative shear 
anisotropy in the early stage of the universe. However, for the case I both 
these parameters vanish. Thus for case I, the universe has no relative shear 
anisotropy. 
\subsubsection{\textbf{Density parameter of matter}}
\begin{figure}[!htb]
\centerline{
  \includegraphics[scale = 0.35]{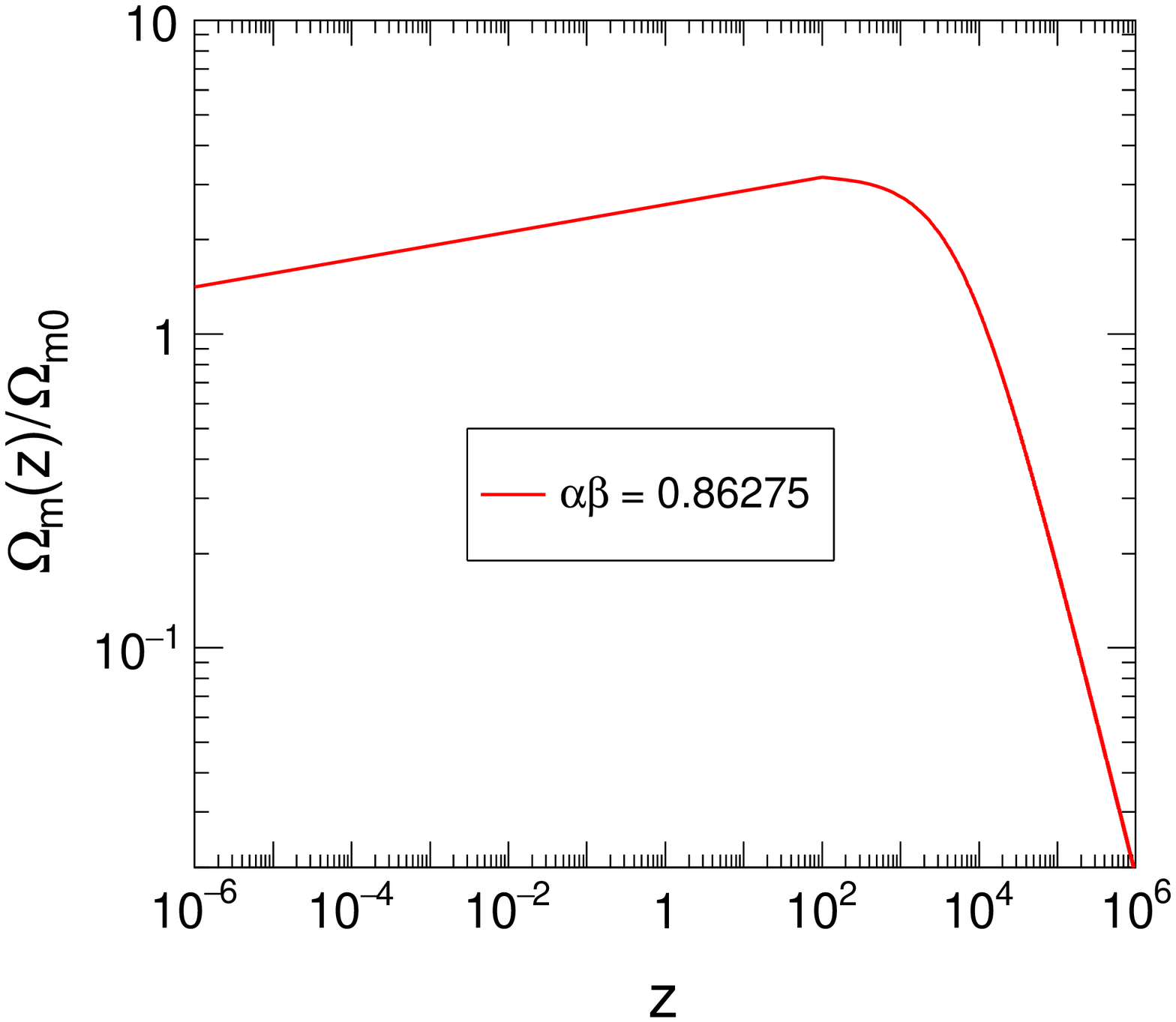}\hspace{1cm} 
  \includegraphics[scale = 0.35]{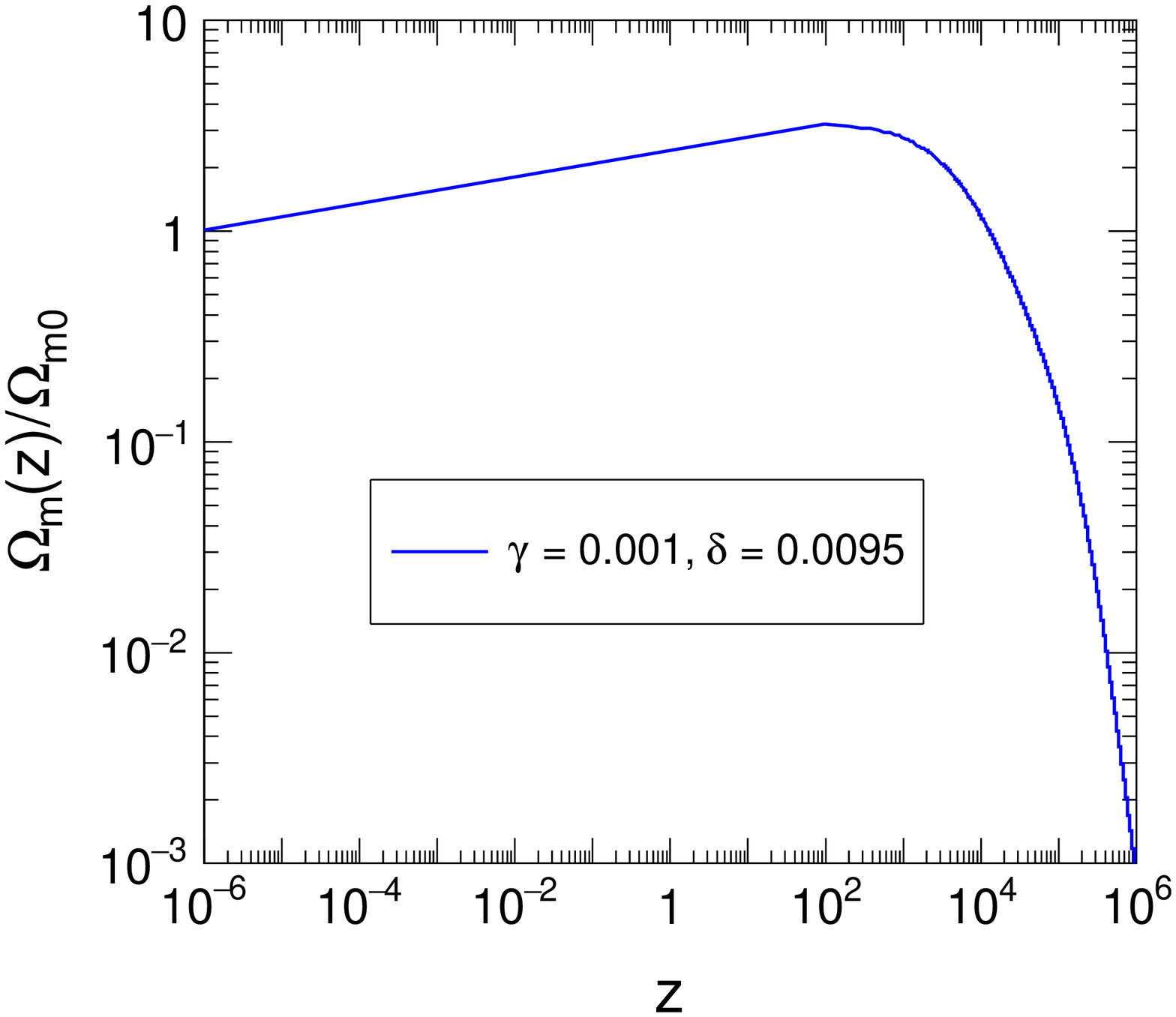}} 
\caption{Variation of matter density parameter with respect to redshift for 
case I (left) and case II (right) with the constrained set of free model 
parameters.}
\label{fig9}
\end{figure}

To visualize how the matter density parameter varies from early stages to 
present epoch according to our anisotropic model of the universe, we have 
plotted the $\Omega_{m}/\Omega_{mo}$ against the redshift in Fig.\ \ref{fig9}
by using Eq.\ (\ref{dens_mat_I}) and (\ref{dens_mat_II}) for the case I and 
case II respectively for the constrained values of free parameters for each 
cases. The left panel of Fig.\ \ref{fig9} is for case I and the right 
panel is for case II. We have plotted this parameter for both the cases 
against the value of $z$ up to $10^{6}$. From the figure, we have seen that 
there is a matter dominated era up to $z$ $\sim 10^{3}$ for both the cases and 
it supports the matter dominated region $0.5 \leq z \leq 3000$ as mentioned in 
Ref.\ \cite{Frieman_2008}. For very high values of $z$, the density parameter 
for matter drops down considerably close to zero. It suggests that the universe 
was radiation dominated at it's early stage. Thus this anisotropic model of the 
universe agrees well with the observed or predicted evolution nature of the 
standard cosmology. However, there are differences between the matter density 
parameter evolution in case I and case II. Firstly, falling of this 
parameter for very high $z$ values is not smooth in the case II in comparison 
to in the case I. Secondly, in the case I the ratio $\Omega_{m}/\Omega_{mo}$ 
moves gradually towards one as $z$ tends to zero as expected, but in the case 
II this ratio reached to the value one before $z$ reached its zero value.   

\subsubsection{\textbf{Age of the universe}}
According to the standard model of cosmology ($\Lambda$CDM) the age of the 
universe is calculated to be $\sim$ 13.786 Gyr. To find the age of the universe 
in our model we have calculated it by using Eqs.\ \eqref{age_caseI} and
\eqref{age_univ_caseII} for the case I and case II respectively with the 
constrained sets of corresponding free parameters. We found that according to 
case I the age of the universe is $\sim$ 13.084 Gyr and according to case II 
it is $\sim$ 13.741 Gyr. Thus we see that case II result agrees very well with 
the results of the standard model of cosmology with a deviation of only 0.33\%. 
Whereas the deviation of the result in case I is 5.09\%.    

\section{Conclusions}\label{5}
In this study we have considered the simplest Bianchi type I metric with slight 
modifications in directional scale factors for two different cases to understand
the possible anisotropy that may exist in the universe. Here, we have used the 
generalized equations of Bianchi type I cosmological model and implemented them
in both the considered cases. Starting from the field equations, we have 
construct all the cosmological parameters including average Hubble parameter, 
luminosity distance, distance modulus, deceleration parameter, equation of 
state, shear scalar and various density parameters etc.\ for both the cases. 
After the mathematical constructions, we have tried to plot various parameters 
with respect to cosmological redshift and compare the results with 
$\Lambda$CDM model which is the best fitted model to cosmological data till 
date. Also we compare the results with various observational data like HKP 
data, SVJ05 data, SJVKS10 data and GCH09 data for the Hubble parameter and with 
Union 2.1 data for the distance modulus. Further, we have used observational 
data of Hubble parameter given in Table \ref{tableI} that are obtained 
from various sources as mentioned in the table to constrained the free 
parameters of both the cases i.e. $\alpha$ and $\beta$ for the case I, and 
$\delta$ and $\gamma$ for the case II as shown in Fig.\ref{fig2}. 
{In this context, explicitly we should state that we have used the 
observational Hubble parameter data to find out the best possible values of 
model parameters as mentioned above. For the case I, we have equated 
Eq.}~\eqref{H_Z_caseI} {with $H_0$ for $z=0$ using the Planck $2018$ data} 
\cite{Planck_2018} {for the same and found that $\alpha \beta = 0.86275$. 
Thus any set of values of $\alpha$ 
and $\beta$ which satisfies this condition are eligible for consideration in 
the case I. In our work, we have not considered any special set of values of 
$\alpha$ and $\beta$ but taken the condition as a whole for the graphical 
analysis. However, for case II this method is not suitable as the 
expression for the Hubble parameter in this case is complicated as compared to 
the case I. Therefore we have fitted the 43 observational Hubble parameter 
data in Table} \ref{tableI} {using the least square fitting technique. 
Then we have changes our model parameters $\delta$ and $\gamma$ in 
Eq.}~\eqref{H_Z_caseII} {for $H(z)$ of the case II to find out which 
values of these parameters give the suitable or close result with the least 
square fitting curve. In our analysis we have found that for $\delta = 0.001$ 
and $\gamma = 0.0095$, the plot of the Hubble parameter fairly agrees with the 
fitting curve as shown in Fig.} \ref{fig2}. {Thus from our analysis we have 
taken  $\alpha\beta = 0.86275$ for the case I, and  $\delta = 0.001$, 
$\gamma = 0.0095$ for the case II as the constrained values of the model 
parameters in all other graphical analysis.} 

After {constraining the model parameters} we have plotted various 
cosmological parameters against cosmological redshift for both the cases with 
the constrained values of free parameters and tried to explain those plots. 
From the results of this analysis we have found that for the case I, where the 
directional scale factors are modified by multiplicative constants, the model 
unable to show any kind of anisotropy as the shear scalars and density 
parameter of anisotropy are zero in this case. It is to be noted that for 
$\alpha = \beta = 1$, the case I is reduced to the $\Lambda$CDM model.
For the case II, we found  {relatively higher amount of anisotropy  as compared to case I} for higher values 
of redshift and it becomes very very small at $z \le$ $20$ (see 
Fig.\ \ref{fig7}). The case II also shows the presence of relative shear 
anisotropy, which is 
obviously very small for small value of redshift. The case II also reduces to 
the $\Lambda$CDM model when $\delta = \gamma = 0$. Moreover, in terms of age 
of the universe the case II is found to be in very good agreement with the 
$\Lambda$CDM model. {Further, we have found that the current value of the 
Hubble parameter ($H_0$) for the case I is $69.31$ and for the case II is 
$67.56$. The result obtained for case II shows more consistency with the 
Planck 2018 result} \cite{Planck_2018}{. Similarly, the current value 
of deceleration parameter for case I is $-0.47$ and for case II is $-0.54$. 
Here also the result obtained from case II is more consistent with the Planck 
$2018$ data} \cite{Planck_2018}{. Furthermore, the constrained values of 
model parameters for both the cases give the matter dominated region in the 
density parameter of matter vs redshift ($z$) plots of Fig.}~\ref{fig9}{, 
which show good agreement within the range $0.5 \leq z \leq 3000$ as 
mentioned in Ref.} \cite{Frieman_2008}{. For the range of values of 
density parameters 
of anisotropy $\Omega_{\sigma 0}$ as suggested in Ref.}~\cite{Akarsu_2019} 
 {i.e.~$10^{-23} \leq \Omega_{\sigma 0} \leq 10^{-15}$, the values of 
$\delta$ and $\gamma$ lies in between $10^{-8}$ and $10^{-3}$ for the case II.}
 
Thus from our work, we have shown that even there exist variations in 
directional scale factors, the Bianchi type I model sometime gives purely 
isotropic universe just like the case I. Also we have found that although 
very small variations of directional scale factors unable to change 
the isotropic nature of the universe significantly in the current state, but it 
may possess {relatively higher} contribution of anisotropy in early stage of the 
universe {compared to the current state}. 

{Finally, it needs to be mentioned that in most of the past studies on 
various anisotropic cosmological models authors tried to constraint the 
cosmological parameters related to the possible anisotropy of the universe 
using the available cosmological data with various statistical methods. For 
example, in Refs.}~\cite{Akarsu_2019, Andrade_2018} {Bayesian 
inference 
technique is used, goodness of fit is used in Refs.}~\cite{Bunn_1996} {and
simulation techniques like Monte Carlo is used in Refs.}~\cite{Hansen_2004, 
Bunn_1996}{. Whereas in our work we constrain our model parameters using 
the available cosmological data as well as the constrained anisotropic density 
parameter to predict the possible anisotropy present in the universe as 
mentioned above. Apart from this difference, our work is mainly based on 
analytical techniques in contrast to statistical methods of previous studies
as mentioned. We have derived analytical expressions of all the cosmological 
parameters based on our model, that contain the anisotropic parameters of the 
model. So these cosmological parameters and hence our findings can hopefully 
be tested with the early universe cosmological data that may be available in 
future from the future advanced telescopes, such as the Thirty Meter 
Telescope} \cite{TMT}{, Extremely Large Telescope} \cite{ELT}{, CTA} 
\cite{CTA} 
{etc.}

\bibliographystyle{apsrev}
\end{document}